\documentclass[aip,jcp,reprint,amsmath,amssymb]{revtex4-2}
\usepackage{dcolumn, bm}
\usepackage{graphicx, tikz, orcidlink}
\usepackage{epstopdf}
\usepackage{ascmac}
\usepackage[normalem]{ulem}
\usepackage{cancel}
\usepackage{here}
\usepackage{multirow}
\usepackage{comment}

\usepackage[]{hyperref}
\usepackage{xcolor}
\definecolor{AIPBlue}{RGB}{61, 180, 229}
\hypersetup{
colorlinks,
linkcolor=AIPBlue,
citecolor=AIPBlue,
urlcolor=AIPBlue
}

\begin{document}
\title{Thermodynamic quantum Fokker--Planck equations and their application to thermostatic Stirling engine}
\date{Last updated: \today}

\author{Shoki Koyanagi \orcidlink{0000-0002-8607-1699}}\email[Author to whom correspondence should be addressed: ]{koyanagi.syoki.36z@st.kyoto-u.jp and tanimura.yoshitaka.5w@kyoto-u.jp}\affiliation{Department of Chemistry, Graduate School of Science,
Kyoto University, Kyoto 606-8502, Japan}

\author{Yoshitaka Tanimura \orcidlink{0000-0002-7913-054X}}
\email[Author to whom correspondence should be addressed: ]{koyanagi.syoki.36z@st.kyoto-u.jp and tanimura.yoshitaka.5w@kyoto-u.jp}
\affiliation{Department of Chemistry, Graduate School of Science,
Kyoto University, Kyoto 606-8502, Japan}

\begin{abstract}
We developed a computer code for the thermodynamic quantum Fokker--Planck equations (T-QFPE), derived from a thermodynamic system-bath model. This model consists of an anharmonic subsystem coupled to multiple Ohmic baths at different temperatures, which are connected to or disconnected from the subsystem as a function of time. The code numerically integrates the T-QFPE and their classical expression to simulate isothermal, isentropic, thermostatic, and entropic processes in both quantum and classical cases. The accuracy of the results was verified by comparing the analytical solutions of the Brownian oscillator. Additionally, we illustrated a breakdown of the Markovian Lindblad-master equation in the pure quantum regime. As a demonstration, we simulated a thermostatic Stirling engine 
 employed to develop non-equilibrium thermodynamics [S. Koyanagi and Y. Tanimura, J. Chem. Phys 161, 114113 (2024)] under quasi-static conditions. The quasi-static thermodynamic potentials, described as intensive and extensive variables, were depicted as work diagrams. In the classical case, the work done by the external field is independent of the system-bath coupling strength. In contrast, in the quantum case, the work decreases as the coupling strength increases due to quantum entanglement between the subsystem and bath. The codes were developed for multicore processors using Open multiprocessing (OpenMP) and for graphics processing units (GPU) using the Compute United Device Architecture (CUDA). These codes are provided as supplementary materials.
\end{abstract}
\maketitle

\section{Introduction}
\label{sec.intro}

In contrast to classical thermodynamics, which is a macroscopic qualitative theory, quantum thermodynamics, described by the Hamiltonian, allows for dynamic investigations in addition to equilibrium cases. A commonly used model for describing thermodynamics is a system--bath (SB) model, which consists of a subsystem interacting with a harmonic heat bath.\cite{GRABERT1988115,Weiss2012,T06JPSJ} Due to quantum entanglement between the system and the bath (bathentanglement),\cite{T20JCP} the quantum SB system exhibits unique properties compared to the classical one, particularly at low temperatures. These include the negativity of the noise correlation function\cite{T06JPSJ} and deviations from the canonical equilibrium state.\cite{GRABERT1988115,Weiss2012,T06JPSJ,T20JCP} To simulate SB entanglement dynamics even under external driving forces, it is necessary to consider both energy relaxation and excitation from the bath, which is related by the fluctuation-dissipation theorem,\cite{TK89JPSJ1} such as for the reduced density operator based-approaches as the hierarchical equations of motion (HEOM) approach,\cite{TK89JPSJ1,T90PRA,IT05JPSJ,T06JPSJ,T20JCP,T14JCP,T15JCP,TW91PRA,TW92JCP,KT13JPCB,ST11JPCA,T15JCP,IT19JCTC,PhysRevB.95.064308,YiJing2022,Petruccione2023,Strunz2024,Thoss2024} and
the quasi-adiabatic path integral (QUAPI) approach.\cite{Makri95, Makri96, Makri96B,Thorwart00,JadhaoMakri08,MakriJCP2014,Reichman2010} Note that the detailed balance and positivity conditions often assumed for the reduced description of the density operator are sufficient conditions resulting from an accurate description of the dynamics, but they are not necessary conditions. In quantum thermodynamics, which deals with the delicate energy balance between the subsystem and the bath, arguments based solely on sufficient conditions may lead to artifacts, such as breaking the second law of thermodynamics, by ignoring bathentanglement.

Among these, the HEOM are stable as long as a sufficient number of hierarchy elements are included and reliable, as their numerical accuracy has been confirmed through comparisons with analytical solutions.\cite{T15JCP}  This approach is suitable for thermodynamic investigations because it allows for numerically rigorous dynamic simulations and evaluates the SB interaction and bath parts of the energy, which cannot be assessed using the conventional reduced equation of motion approach, even for processes far from equilibrium.\cite{KT15JCP,KT16JCP,ST20JCP,ST21JPSJ,KT22JCP1,KT22JCP2,KT24JCP1,KT24JCP2}

It should be noted that there are two types of non-Markovian effects: one originating from the noise correlation time determined by the bath spectral distribution function (SDF) and the other from the bath temperature determined by the Matsubara frequency. The former is important for ultrafast vibrational, electronic, and excitonic processes in the condensed phase, where the subsystem's characteristic time scale and the environment's noise correlation time are both in the femtosecond to picosecond range. However, this is not essential in thermodynamic studies where the subsystem changes on the nanosecond scale or less. In contrast, the latter produces effects unique to quantum thermodynamics, such as changing the equilibrium distribution from a canonical one due to bathentanglement, even in the case of the Ohmic SDF without a cutoff.\cite{IT19JCTC} 

Thus, the low-temperature quantum Fokker--Planck equations (LT-QFPE),\cite{IT19JCTC} were derived from the Brownian (or Ullersma--Caldeira--Leggett) model\cite{Ullersma1966_1,Ullersma1966_2,CALDEIRA1983587} for the Ohmic SDF, include the low-temperature Matsubara terms in the hierarchical form without resorting to the rotating wave and factorized approximations. The LT-QFPE is useful for thermodynamic exploration because they reduce the numerical simulation cost compared to the regular HEOM for the Drude SDF\cite{TW91PRA,TW92JCP,KT13JPCB,ST11JPCA,T15JCP} and makes it possible to take the classical limit to obtain the Kramers equation, which is equivalent to the Langevin equation used for thermodynamic simulations in molecular dynamics (MD) approaches. The source code of the LT-QFPE for simulating single and multi-state Brownian systems is presented as a supporting material in Ref. \onlinecite{IT19JCTC}, along with a demo program comparing LT-QFPE results for non-adiabatic dynamic simulations with those of the fewest-switch surface hopping and Ehrenfest methods with a classical Markovian Langevin force. The results reveal the importance of the quantum low-temperature correction terms.

To construct a systematic theory of thermodynamics, it is essential to include a thermostatic process.\cite{KT24JCP1} By introducing multiple baths at different temperatures, we can study a thermostatic process in which temperature varies with time. Assuming that the time scale of quantum thermal fluctuations is shorter than the time scale of the external perturbations, we can extend LT-QFPE to the thermostatic case by introducing time-dependent Matsubara frequencies, described as thermodynamic quantum Fokker--Planck equations  (T-QFPE).\cite{KT24JCP1,KT24JCP2}  This article presents an overview of the T-QFPE, provides the C++ source code for numerical simulations, and briefly explains its theoretical background to further the development of quantum thermodynamics. 

In Section \ref{T-QFPE}, after explaining the thermodynamic SB Hamiltonian, we present the T-QFPE in the Wigner representation. 
In Section \ref{sec:QSE}, we verify the accuracy of the T-QFPE code by comparing it with analytical solutions of the Brownian oscillator. 
In Section \ref{CQStirling}, we then demonstrate the capability of our code by simulating the quasi-static processes of the thermostatic Stirling engine, presenting the work diagram in terms of intensive and extensive variables. Section \ref{sec:conclude} presents concluding remarks. 

\section{Thermodynamic Quantum Fokker--Planck equations (T-QFPE)}
\label{T-QFPE}

The T-QFPE were developed for Ohmic SDF to investigate various thermodynamic processes, including isothermal, thermostatic, isentropic, thermostatic, and entropic processes.\cite{KT24JCP1, KT24JCP2} For these processes, the free energy can also be evaluated as a function of thermodynamic intensive and extensive variables by assessing the work done for changes in temperature and external fields. The model Hamiltonian of T-QFPE is expressed as follows:
\begin{eqnarray}
\label{eq:Htot}
\hat{H}_{\rm tot} ( t ) = \hat{H}_{\rm A} ( t ) + \sum_{k = 1}^N \hat{H}_{\rm IB}^k ( t ) ,
\end{eqnarray}
where $\hat{H}_{\rm A} ( t )$ is the subsystem Hamiltonian. Here, we consider the case where the subsystem is described in phase space as:
\begin{eqnarray}
\label{eq:HA}
\hat{H}_{\rm A} ( t ) = \frac{\hat{p}^2}{2 m} + U ( \hat{q} , t ) ,
\end{eqnarray}
where $\hat{q} , \hat{p} , m$ are the position and momentum operators, and the particle mass, respectively, and $U ( \hat{q} , t )$ is the potential. To treat thermostatic processes, in Eq. \eqref{eq:Htot}, we consider $N$ heat baths with the different temperatures $T_k$. The $k$th bath Hamiltonian including the system--bath (SB) interaction between the subsystem and the $k$th bath $\hat{H}_{\rm IB}^k ( t )$ is expressed as:
\begin{eqnarray}
\label{eq:HIB}
\hat{H}_{\rm IB}^k ( t ) = \sum_j \left\{ \frac{ ( \hat{p}_j^k )^2 }{2 m_j^k } 
+ \frac{m_j^k ( \omega_j^k )^2 }{2}  \left[ \hat{x}_j^k - \frac{A_k \xi_k ( t ) c_j^k \hat{q}}{m_j^k ( \omega_j^k )^2} 
\right]^2 \right\} , \nonumber \\
\end{eqnarray}
where $\hat{x}_j^k , \hat{p}_j^k , m_j^k , \omega_j^k ,$ and $c_j^k$ are the position and momentum operators, mass, angular frequency, and coupling coefficient of the $j$th mode in the $k$th bath, respectively, and $A_k$ is the SB coupling strength between the subsystem and the $k$th bath. Here, $\xi_k ( t )$ is the window function that is $1$ when the bath is attached to the subsystem and $0$ when it is not. 

For the $k$th bath, the subsystem A is driven by the external force $\hat X^k (t)$ through the interaction $-V(\hat q) \hat X^k$,
where $\hat{X}^k(t)$ 
is the Heisenberg representation of $\hat{X}^k \equiv \sum_j c_j^k \hat x_j^k$  for $\hat H_{\rm B}^k$. The character of the bath is determined by SDF of the $k$th bath, defined as $J_k ( \omega ) = \sum_j ({A_k^2 ( c_j^k )^2}/{2 m_j^k \omega_j^k}) \delta ( \omega - \omega_j^k ) $. 
Each bath being harmonic and Gaussian in nature, the character of 
$\hat X^k (t)$ is specified by its two-time correlation functions, such as
the symmetrized and canonical correlation functions defined by $C^k(t) \equiv
\langle
\hat{X}^k(t)\hat{X}^k(0)+\hat{X}^k(0)\hat{X}^k(t)\rangle_\mathrm{B}/2$
and 
$R^k(t) \equiv \int_0^{\beta_k} {d\lambda } \, \left\langle {\hat X}^k( - {\rm i}\hbar \lambda )\hat X^k(t ) \right\rangle_\mathrm{B}/2$,
where $\langle \cdots \rangle_\mathrm{B}$ represents the thermal average of the $k$th bath degree of freedom.\cite{T06JPSJ,T15JCP}

The SDF of the $k$th bath is defined as $J_k ( \omega ) = \sum_j ({A_k^2 ( c_j^k )^2}/{2 m_j^k \omega_j^k}) \delta ( \omega - \omega_j^k ) $ and we assume that it is the Ohmic SDF expressed as:
\begin{eqnarray}
\label{eq:SDF}
J_k ( \omega ) = \frac{A_k^2}{\pi} \hbar \omega. 
\end{eqnarray}  
In the classical case, this SDF describes Markovian dynamics. 
The correlation functions are then evaluated as:\cite{IT19JCTC} 
\begin{eqnarray}
\label{eq:sym-approx0}
\begin{split}
C^k ( t ) & \simeq \frac{2 A_k^2 }{\beta_k} \left( 1 + \sum _{l = 1}^{K} 2 \right) \delta (t) 
-\sum _{l = 1}^{K} \frac{2 A_k^2 \nu _{l}^k}{\beta_k} e^{-\nu _{l}^k | t |} 
\end{split}
\end{eqnarray}
and
\begin{eqnarray}
\label{eq:rlx-approx0}
R^k ( t ) = A_k^2 \delta (t).
\end{eqnarray}

We consider the case in which the subsystem attaches to only one bath at the same time; thus the SB coupling strength and temperature of the bath attached to the subsystem are expressed as:
\begin{eqnarray}
A ( t ) = \sum_{k = 1}^N A_k \xi_k ( t ) ,
\end{eqnarray}
and
\begin{eqnarray}
\label{eq:BathT}
T ( t ) = \sum_{k = 1}^N T_k \xi_k ( t ) ,
\end{eqnarray}
respectively. The inverse temperature of the bath is defined as $\beta ( t ) = 1 / k_{\rm B} T ( t )$.  
The window function (thermostatic field,) is defined as:
\begin{eqnarray}
\label{eq:xi}
 \xi_k (t)= \theta(t-t_k)\theta(t_k +\Delta t -t),
\end{eqnarray}
where  $\theta (t)$ is the step function, and the time $t_k$ is defined as $t_k = t_0 + ( k - 1) \Delta t$, with  initial time $t_0$ and  duration $\Delta t$.

For $N$ heat baths, the HEOM consists of a $(K \times N$)-dimensional hierarchy.\cite{KT15JCP,KT16JCP,KT22JCP2} 
We consider the situation where the difference between the inverse temperatures of successive heat baths, for example, $k$ and $k+1$, denoted as $\Delta \beta$, and the time duration of each heat bath $\Delta t$ is expressed as $\hbar \Delta \beta /\Delta t \ll 1$, $\beta ( t )$ [or $T ( t ) $] is considered to change continuously.
Under these conditions, the SB coherence among different heat baths [e.g., the $k$th bath and the $(k+1)$th bath] arising from the bathentanglement between different heat baths becomes negligible, while it is still taken into account in the equilibrium and non-equilibrium distributions of the subsystem.\cite{KT24JCP4}

This allows us to extend the low-temperature quantum Fokker--Planck equations (LT-QFPE)\cite{IT19JCTC} for the set of Wigner distribution functions (WDFs), $W_{\vec{n}}(p, q; t)$, where $\vec{n} = ( n_1 , \cdots , n_K )$ is the nonnegative integer vector that serves as the index of the hierarchy members by introducing time-dependent Matsubara frequencies $\nu(t) = 1/\hbar \beta(t)$. 

Thus, the T-QFPE can be expressed as follows:\cite{KT24JCP1,KT24JCP2}
\begin{eqnarray}
\label{eq:T-QFPE}
    &&\frac{\partial  {W}_{\vec{n}} (p,q; t) }{\partial t} \nonumber \\
    &&=-\left(\mathcal{\hat L}_{qm}(t)+\sum _{l = 1}^{K}n_{l}\nu_{l}(t)+\hat{\Xi }_{K} (p,q; t)\right) {W}_{\vec{n}} (p,q; t) \nonumber \\
    &&\quad -\sum _{l=1}^{K}\hat{\Phi }_p(t) {W}_{\vec{n}+\vec{e}_{l}} (p,q; t) \nonumber \\
    &&\quad -\sum _{l=1}^{K}n_{l}\nu _{l}(t)\hat{\Theta }_{l}(p,q; t) {W}_{\vec{n}-\vec{e}_{l}} (p,q; t),
\end{eqnarray}
where $\vec{e}_l$ is the unit vector, whose $l$th element is $1$, otherwise $0$. The frequency $\nu_l ( t )$ is the $l$th characteristic frequency of the non-Markovian effect arising from the Matsubara frequency, and $K$ is the number of frequencies.\cite{T90PRA,IT05JPSJ,T14JCP}  Although the physical meaning is no longer clear, an efficient and simple approach to suppressing $K$ is to use the Pad{\'e decomposition represented by $\nu_l ( t ) = \zeta_l k_{\rm B} T ( t ) / \hbar$, where $\zeta_l$ is the $l$th Pad{\'e} factor for frequency.\cite{YanPade10A,YanPade10B,YanPade11}  

In general, the quantum correlations described by the Matsubara frequencies decay quickly and the above equations describe the thermostatic process accurately.
However, for a Brownian system, when the correlation time exceeds the time scale of the bath temperature change, the SDF-based description of a thermal bath breaks down. This situation arises when $A ( t )$ changes faster than the noise correlation time, breaking the translational symmetry of the reduced equations of motion regarding the position (see Appendix \ref{sec:AdditionalTerm}).
In such cases, results must be verified by comparing them with those explicitly including multiple heat baths using the $(K \times N$)-dimensional hierarchy,\cite{KT24JCP4} as has been done for heat transfer processes\cite{KT15JCP,KT16JCP} and for the adiabatic processes of quantum Carnot cycles.\cite{KT22JCP2}

The operator $\hat{\mathcal{L}}_{qm} ( t )$ is the quantum Liouvillian of the subsystem expressed using the Fourier form of the potential expressed as:\cite{T15JCP,KT13JPCB,TW91PRA,TW92JCP,ST11JPCA,ST13JPSJ,ST14NJP,GST16JPSJ,IDT19JCP,Frensley1990}
\begin{eqnarray}
\label{eq:QMLiouvillian2}
\begin{split}
-\mathcal{\hat L}_{qm} ( t ) &W ( p , \, q ) \equiv
- \frac{p}{m} \frac{\partial W ( p , \, q) }{\partial q} \\
&- \frac{1}{\hbar} \int_{-\infty}^{\infty} \frac{d p'}{2 \pi \hbar} U_{\rm W} ( p - p' , q ; t)
W( p' , \, q ) ,
\end{split}
\end{eqnarray}
where
\begin{eqnarray}
 \quad U_{\rm W} ( p , q ;t ) &&= 2 \int_0^\infty d x \sin \left( \frac{p x}{\hbar} \right) \nonumber \\
&& \times \left[ U \left( q + \frac{x}{2}; t \right) - U \left( q - \frac{x}{2} ; t \right) \right].
\end{eqnarray}
While the above representation of Liouvillian is numerically stable even in cases where the potential is singular (the source code for the above scheme is given in supporting material of Ref. \onlinecite{T15JCP}), when the potential is in a simple analytical form, we employ the Moyal expansion:\cite{IT19JCTC,Frensley1990,risken1996fokker}
\begin{eqnarray}
\label{eq:QMLiouvillian}
&-& \mathcal{\hat L}_{qm}(t)W (p,\,q) \equiv - \frac{p}{m}\frac{\partial W (p,\,q)}{\partial q} \nonumber \\
&+& \sum_{n = 0}^\infty \frac{1}{( 2 n + 1 ) !} \frac{\partial^{2 n + 1} U(q; t )}{\partial q^{2 n + 1}}
\left( - \frac{\hbar^2}{4} \frac{\partial^2}{\partial p^2} \right)^n \frac{\partial W ( p , q )}{\partial p}
.\nonumber \\
\end{eqnarray}
The other operators in Eq. \eqref{eq:T-QFPE} are defined as follows:\cite{KT24JCP1,KT24JCP2},
\begin{eqnarray}
    \hat{\Phi}_p (t) \equiv - \frac{A ( t )}{\beta ( t )} \frac{\partial }{\partial p},
\end{eqnarray}
\begin{eqnarray}
\label{Theta}
      \hat{\Theta }_{0}(p,q; t) =\frac{A ( t ) \beta ( t )}{m} \biggl(p+\frac{m}{\beta(t) }\frac{\partial }{\partial p}\biggr),
\end{eqnarray}
\begin{eqnarray}
\label{Thetal}
    \hat{\Theta }_{l}(p,q; t)\equiv 2 A ( t ) \eta_l    \frac{\partial }{\partial p},
\end{eqnarray}
for $1\le l \le K$, and
\begin{eqnarray}
      \hat{\Xi }_{K} (p,q; t)\equiv \hat{\Phi}_p (t) \sum _{l = 0}^{K}\hat{\Theta }_{l} ( p , q ; t ) ,
\label{Xipq}
\end{eqnarray}
where $\eta_l$ is the $l$th Pad{\'e} coefficient. Note that while various numerical techniques have been developed to fit the bath correlation function [Eq. \eqref{eq:sym-approx0}], \cite{10.1063/5.0095961,PhysRevLett.129.230601} here we chose the Pad{\'e} spectral decomposition because $\zeta_l$ and $\eta_l$ are independent of the change in bath temperature over time under the thermostatic process. The factors $\zeta_l$ and $\eta_l$ are listed in Ref. \onlinecite{IT19JCTC}. In the numerical simulation, we truncate $W_{\vec{n}} ( p , q ; t )$ satisfying the condition $\sum_{j = 1}^K n_j \leq N$, where $N$ is the truncation integer.\cite{IT05JPSJ,T14JCP,T15JCP,T20JCP} 

The classical limit of T-QFPE is the Kramers equation expressed as:\cite{TW91PRA,TW92JCP,KT13JPCB,ST11JPCA,KRAMERS1940284,risken1996fokker}
\begin{eqnarray}
\frac{\partial {W} (p, q ; t) }{\partial t} &=& -\mathcal{\hat L}_{cl} ( t ) W (p , q ; t) \nonumber \\ 
 &+& \frac{A^2}{m}  \frac{\partial }{\partial p} \left( p + \frac{m}{\beta (t)} \frac{\partial }{\partial p} \right) W(p , q ;t), \nonumber \\
\label{heom_cl}
\end{eqnarray}
where $W ( p , q )$ is the phase distribution function and the classical Liouvillian is defined as:
\begin{eqnarray}
-\mathcal{\hat L}_{cl}(t)W(p , q) \equiv -\frac{p}{m}\frac{\partial W(p , q)}{\partial q}  + \frac{\partial U(q; t)}{\partial q}\frac{\partial W(p , q)}{\partial p}.\nonumber \\
\label{L_cl}
\end{eqnarray}
The description of the Kramers equation is equivalent to that of the Langevin equation.\cite{TW91PRA,T06JPSJ} 

Note that the Wigner representation is also convenient for implementing various boundary conditions, most notably periodic,\cite{TW92JCP,KT13JPCB,IDT19JCP} open,\cite{TM97JCP} and inflow--outflow boundary conditions.\cite{ST13JPSJ,ST14NJP,GST16JPSJ}

As presented in Refs. \onlinecite{KT24JCP1, KT24JCP2}, thermodynamic intensive and extensive variables, which are interrelated by time-dependent Legendre transformations, and various thermodynamic potentials can be evaluated from the T-QFPE. For the potential expressed as $U ( q ; t ) = U_0 ( q ) - x ( t ) q$, where $U_0 ( q )$ is the time-independent part of the potential, the enthalpy of the subsystem $H_{\rm A} ( t )$ is expressed as:
\begin{eqnarray}
\label{eq:Enthalpy}
H_{\rm A} ( t ) = U_{\rm A} ( t ) - x ( t ) X_{\rm A} ( t ) ,
\end{eqnarray}
where 
\begin{eqnarray}
\label{eq:InternalEnergy}
U_{\rm A} ( t ) = \mathrm{tr}_{\rm A} \left\{ \left[ \frac{p^2}{2 m} + U_0 ( q ) \right] W ( p , q ; t ) \right\} 
\end{eqnarray}
is the internal energy and
\begin{eqnarray}
\label{eq:XA}
X_{\rm A} ( t ) = \mathrm{tr}_{\rm A} \{ q W ( p , q ; t ) \}.
\end{eqnarray}
Here, $W ( p , q ; t )$  is the zeroth number of the solution of Eqs. \eqref{eq:T-QFPE} expressed as ${W}_{\vec{0}} (p,q,t)$ in the quantum case and the solution of the thermodynamic Kramers equation \eqref{heom_cl} in the classical case.

Details of numerical simulations with T-QFPE are presented in Appendix \ref{sec:Detail}.  

\section{Numerical examination for isothermal Brownian oscillator system}
\label{sec:QSE}

\begin{figure}
\centering
\includegraphics[width=8cm]{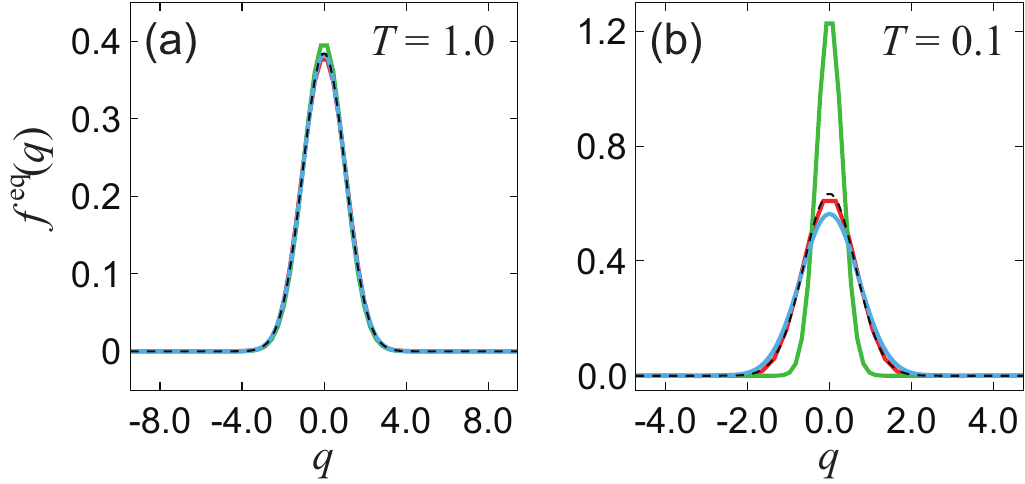}
\caption{\label{fig:Eq} Non-Markovian tests for the description of (i) the thermal equilibrium state $f^{\rm eq}(q)$ at 
 (a) high temperature [$T = 1.0$] and (b) low temperature [$T = 0.1$] calculated for an Ohmic Brownian oscillator system.  In each figure,  the calculated results from the T-QFPE (red curve), Kramers equation (green curve), quantum master equation (blue curve), and analytical solution (black dashed curve) are depicted, respectively. The SB coupling strength is $A = 1.0$, and the frequency of the oscillator is $\omega_0 = 1.0$}
\end{figure}

\begin{figure}
\centering
\includegraphics[width=8cm]{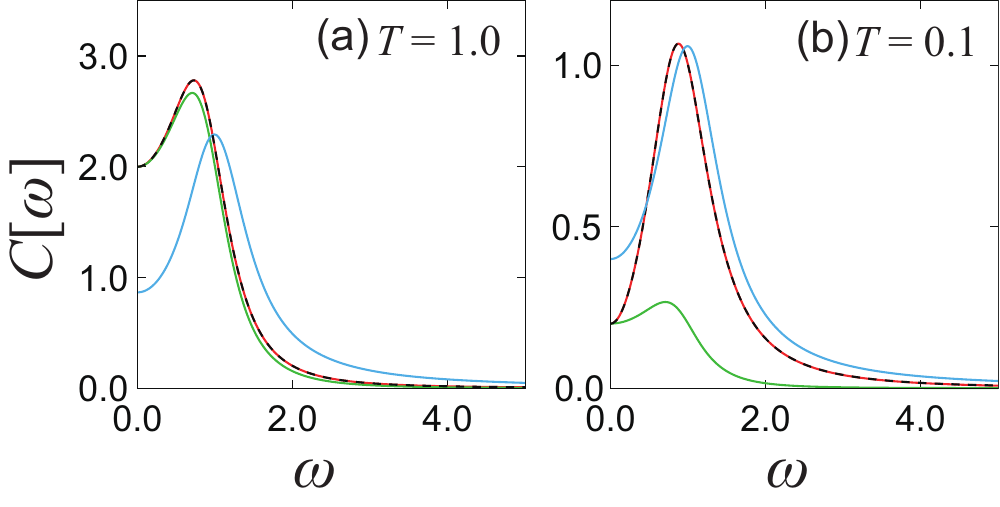}
\caption{\label{fig:CF} Non-Markovian test for (ii) the auto-correlation function $C [\omega]$ to examine the effects of temperature effects. In each figure, the results calculated from the T-QFPE (red curve), Kramers equation (green curve), quantum master equation (blue curve), and analytical solution (black dashed curve) are plotted as a function of $\omega$. The SB coupling strength was $A = 1.0$, and the bath temperatures were (a) $T = 1.0$ (high), and (b) $T = 0.1$ (low).}
\end{figure}

\begin{figure}
\centering
\includegraphics[width=4cm]{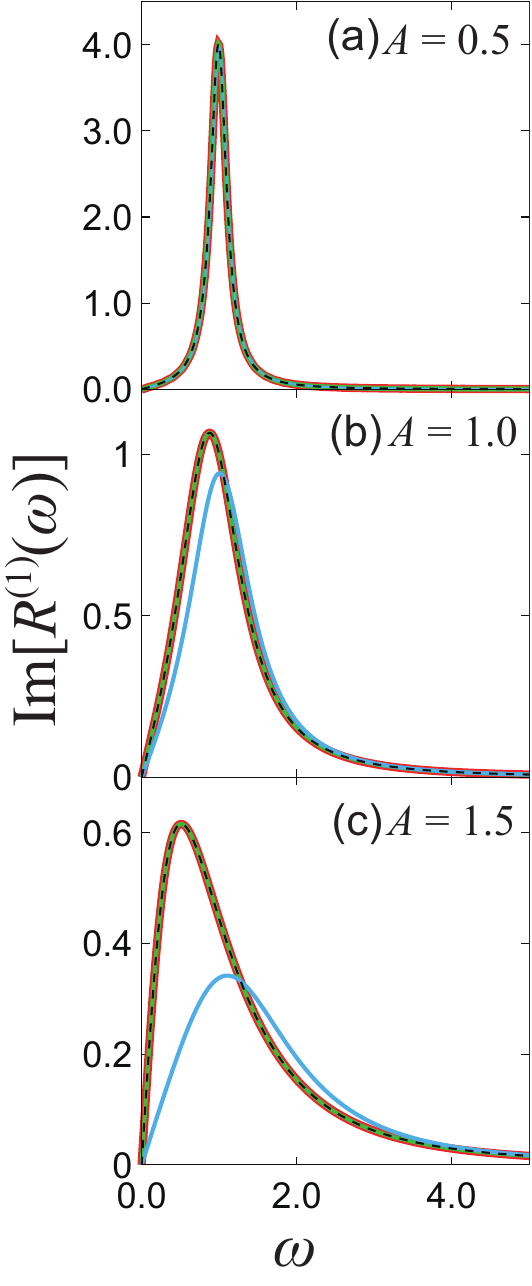}
\caption{\label{fig:LRF} Non-Markovian test on (iii) the linear response function ${\rm Im}[R^{(1)} ( \omega )]$ to investigate the description of non-perturbative SB coupling effects. In each figure, the results computed from the T-QFPE (red curve), the Kramers equation (green curve), the quantum master equation (blue curve), and the analytical solution (black dashed curve) are plotted as a function of $\omega$ in the high temperature case $T = 1.0$.
The SB coupling strengths are (a) $A = 0.5$ (weak), (b) $A = 1.0$ (intermediate), and (c) $A = 1.5$ (strong), respectively.
In the weak coupling case, all results overlapped and were indistinguishable.  In the case of intermediate and strong coupling, the QME results deviated from the exact results, while the T-QFPE and Kramers results always overlapped with the exact results in this harmonic Brownian case where quantum effects do not play a role.}
\end{figure}

Proceeding with a simulation without verifying computational precision is akin to ascending a cliff without the assurance of a rope. Thus, although limited to the case where the subsystem is harmonic, numerically ``exact'' tests (non-Markovian tests) for reduced dynamics of a subsystem under nonperturbative and non-Markovian SB interactions have been developed based on exact analytical solutions of the Brownian oscillator.\cite{T15JCP} Here, non-Markovian effects refer to the effects of time correlations in the noise generated by the heat bath. There are two types of non-Markovianity: one that exists in both the classical and quantum cases, described as the cutoff frequency of SDF, and the other that exists only in the quantum case, described by Matsubara frequency. Since quantum thermodynamic processes are slower than the bath-noise correlation time, we focus here on the latter non-Markovian effects, which arise even from the Ohmic SDF without a cutoff.

The harmonic potential of the subsystem is described as:\cite{GRABERT1988115,Weiss2012}
\begin{eqnarray}
\label{eq:BO}
U ( \hat{q}) = \frac{1}{2} m \omega_0^2 \hat{q}^2,
\end{eqnarray}
where $m$ and $\omega_0$ are the mass and frequency of the subsystem, respectively. In the energy eigenstate representation, the above system is described as $\hat{H}_A= \hbar \omega_0 ({\hat a}^+{ \hat a}^- +1/2)$,
where ${\hat a}^{\pm}$ are the creation--annihilation operators of the eigenstates.

Non-Markovian tests [Ref. \onlinecite{T15JCP} ] are based on the solutions for (i) the steady-state distribution to examine the accuracy of the thermodynamic description, (ii) the symmetric autocorrelation function $C(t)\equiv \langle \hat q(t) \hat q+ \hat q \hat q(t) \rangle/2$ to examine the description of temperature effects, (iii) the linear response function $R^{(1)}(t)\equiv i\langle [\hat q(t), \hat q] \rangle/\hbar$ to examine the description of non-perturbative SB coupling effects, and (iv) the nonlinear response function $R_\mathrm{TTR}^{(2)}(t_2,t_1)=-\langle[[\hat q^2 (t_1+t_2), \hat q(t_1)], \hat q] \rangle/{\hbar^2}$ to test the dynamical correlation between the system and the bath. Here, the equilibrium expectation value is defined as
$\langle \cdot \cdot \cdot \rangle = \mathrm{tr} \{  \cdot  \cdot \cdot \exp({-\beta \hat H_{\rm tot}}) / Z_{\rm tot} \}$, where $Z_{\rm tot}$ is the partition function for the total Hamiltonian $\hat H_{\rm tot}$. Since the effect of (iv) is not important in the Ohmic case, we perform only tests (i)-(iii).

For all our computations, we fixed the oscillator frequency as $\omega_0 = 1.0$. The temperature was set to (a) $T = 1.0$ (high) and (b) $T = 0.1$ (low).  For the T-QFPE, we set the truncation number of the hierarchy to $N = 7$ in tests (i) and (ii), and we set $N = 6, 7$, and $8$ for the weak ($A = 0.5$), intermediate ($A = 1.0$), and strong ($A = 1.5$) SB coupling cases in test (iii), respectively. We set the number of Pad{\'e} frequencies in the T-QFPE to $K = 2$ and $4$ at high ($T = 1.0$) and low ($T = 0.1$) temperatures, respectively. The mesh size of the WDF was set to $n_q = 64$ and $n_p = 64$. For the T-QFPE, we chose the mesh steps of $d q = 0.3$ and $d p = 0.5$, $d q = 0.3$, and $d p = 0.3$ for high ($T = 1.0$) and low ($T = 0.1$) temperatures, respectively. For the Kramers equation, we chose the mesh steps $d q = 0.15$ and $d p = 0.2$, and $d q = 0.3$ and $d p = 0.4$ for high ($T = 1.0$) and low ($T = 0.1$) temperatures, respectively.

To illustrate the applicability of the Markovian assumption (or stochastic thermodynamics), we also present the results obtained from the quantum master equation (QME) in the Lindblad form.\cite{Lindblad1976} Under the factorized initial condition and rotating wave approximation (RWA) that are employed to preserve positivity under non-realistic Markovian assumption,\cite{T06JPSJ,T20JCP} the QME can be expressed as follows:
\begin{eqnarray}
\label{eq:QME}
\begin{split}
\frac{\partial}{\partial t} \hat{\rho}_{\rm A} ( t )
&= - \frac{i}{\hbar} \left[ \hat{H}_{\rm A} ( t ) , \hat{\rho}_{\rm A} ( t ) \right] \\
&
+ \frac{A^2}{2 m} \bar{n} \left( 2 \hat{a}^+ \hat{\rho}_{\rm A} ( t ) \hat{a}^- 
- \{ \hat{a}^- \hat{a}^+ , \hat{\rho}_{\rm A} ( t ) \} \right) \\
&
+ \frac{A^2}{2 m} ( \bar{n} + 1 ) \left( 2 \hat{a}^- \hat{\rho}_{\rm A} ( t ) \hat{a}^+
- \{ \hat{a}^+ \hat{a}^- , \hat{\rho}_{\rm A} ( t ) \} \right) ,
\end{split}
\end{eqnarray}
where $\bar{n} = 1 / (e^{\beta \hbar \omega_0} - 1)$ and we ignored the non-resonant terms, such as $ (\hat{a}^+ )^2 \hat{\rho}_{\rm A} ( t ) $ and $ \hat{a}^+  \hat{\rho}_{\rm A} ( t ) \hat{a}^+ $.\cite{T06JPSJ}  To perform numerical calculations, we employed 16 eigenstates. 

We first examined the descriptions of the thermal equilibrium state as the function of coordinate as\cite{IT19JCTC} 
\begin{align}
  f^{\rm eq}(q)&\equiv \int \!dp\,W^{\rm eq} (p,q).
\end{align}
In the high--temperature case shown in Fig. \ref{fig:Eq}(a), all calculated results show similar profiles because the thermodynamic characteristic of the subsystem is determined by $\beta \hbar \omega_0$ and the high-temperature limit $\beta \rightarrow 0$ is equivalent to the classical limit $\hbar \rightarrow 0$. However, in the low-temperature case shown in Fig. \ref{fig:Eq}(b), the quantum results are more diffuse than the classical result due to zero-point oscillations. The width of the QME result in Fig. \ref{fig:Eq} (b) is wider than the exact and T-QFPE results due to the lack of bathentanglement.

In Fig. \ref{fig:CF}, we tested the effects of fluctuations at different temperatures. While the equilibrium distributions were similar, the dynamical behaviors of the T-QFPE and Kramers results were very different from the QME result, especially in the low-frequency regime. This is because RWA ignores the double excitation and de-excitation described by the operators $\hat{a}^+\hat{a}^+$ and $\hat{a}^- \hat{a}^-$,\cite{T06JPSJ} which play an important role in high-temperature regimes where the thermal excitation energy is large. At low temperatures, the effect of double excitation and de-excitation is suppressed, and the QME and T-QFPE results become closer, while the classical Kramers results become very different due to the lack of zero-point oscillation. The difference between the QME and the T-QFPE results is due to bathentanglement, which becomes apparent at low temperatures.

In Fig. \ref{fig:LRF}, we tested the effects of dissipation at different SB coupling strengths. As the analytical solution of the harmonic Brownian system indicated,\cite{GRABERT1988115,T15JCP} the linear response function does not exhibit any quantum effects and the analytical, T-QFPE, and Kramers results always overlap, while the results obtained from the QME differ from the others, especially for larger $A$, because the QME is a perturbative approach.

These results indicate that the equations of motion derived using the Markovian or rotating wave (or secular) approximation, such as the Lindblad equation and QME, can only be applied to high-temperature regions where the subsystem exhibits semiclassical dynamics.\cite{T06JPSJ,T20JCP,T14JCP,T15JCP}  
Thus, the validity of the fluctuation theorem\cite{RevModPhys.81.1665,RevModPhys.83.771,sagawa20232law} and stochastic thermodynamics\cite{Schmiedl_2008,Stochastic_Seifert_2012,Esposito2011,strasberg2022quantum,RaoEsposit2018,PhysRevE.105.064124} based on a factorized initial condition and/or Markovian assumption, especially in the quantum case, should be carefully examined. Also, as can be seen from the form of Eq.\eqref{eq:T-QFPE}, it is the fluctuation and dissipation terms that drive the subsystem to the thermal equilibrium state,\cite{T15JCP} and the detail balance condition is not necessary.\cite{KT24JCP2}  In fact, the time evolution of the reduced system described by the theory based on the detail balance condition is different from that described by the SB model. Moreover, the HEOM approach, which satisfies the fluctuation-dissipation theorem, can correctly describe the effects of bathentanglement, while theories based on the detailed balance condition ignore them completely.

Before performing a T-QFPE simulation, these tests should be used to select the working parameters, such as mesh size, time step, and hierarchy depth. Since these tests are limited to the case of harmonic potential, a finer choice of mesh and time step is necessary when anharmonicity is large.

\section{Demonstration with thermostatic Stirling engine}
\label{CQStirling}

\begin{figure}
\centering
\includegraphics[width=6cm]{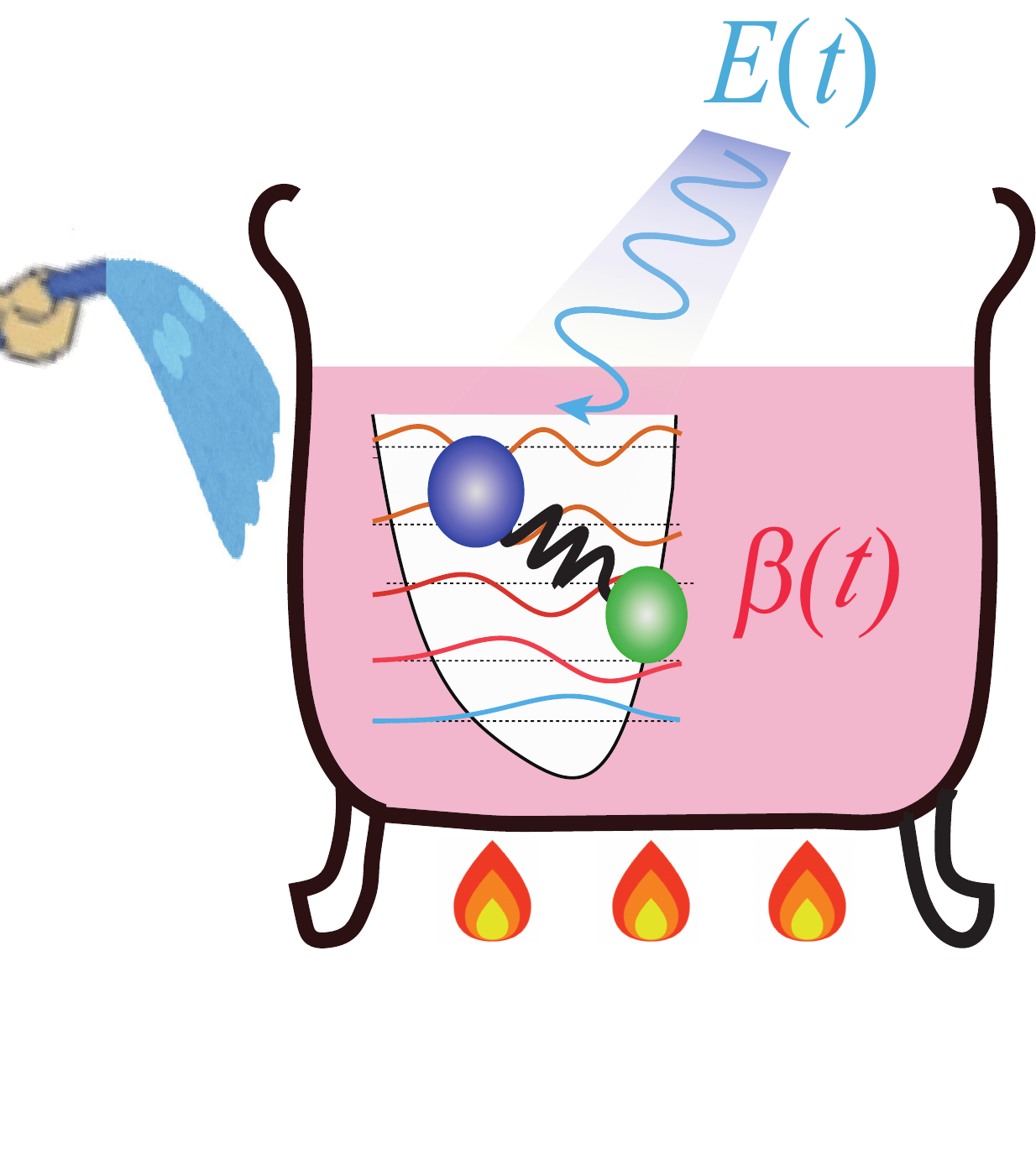}
\caption{\label{fig:Potential}Schematic illustration of a Stirling engine}
\end{figure}

In conventional studies, the quantum Stirling engine employs two isothermal (hot and cold) processes. \cite{wu1998performance,Stirling2021,huang2014quantum,xia2024performance} Here, however, we consider two thermostatic processes, as in the original Stirling engine where the bath temperature changes continuously.\cite{KT24JCP2}
Thus, the thermostatic Stirling engine consists of four steps: (i) a hot isothermal process, (ii) a thermostatic transition from hot to cold, (iii) a cold isothermal process, and (iv) a thermostatic transition from cold to hot, as described by the external field $E( t )$ and the inverse temperature $\beta ( t )$ in Table \ref{Cycelarameter}. The amplitudes are set as $E_1 = 0.5$ and $E_2 = 0.2$, with the inverse temperatures of the hot and cold baths being $\beta_{\rm H} = 1.0$ and $\beta_{\rm C}= 1.5$.
For a subsystem, we consider an anharmonic system, the same as described in Refs. \onlinecite{KT24JCP1} and \onlinecite{KT24JCP2}. The potential $U ( \hat{q} , t )$ is expressed as:
\begin{eqnarray}
\label{eq:Potential}
U ( \hat{q} , t ) = U_2 \hat{q}^2 + U_3 \hat{q}^3 + U_4 \hat{q}^4 - E( t ) \hat{q} ,
\end{eqnarray}
where $U_2 , U_3,$ and $U_4$ are constants, and $E(t)$ is the time-dependent external force. A schematic of the model is presented in Fig. \ref{fig:Potential}. The potential constants are $U_2 = 0.1 , U_3 = 0.02,$ and $U_4 = 0.05$. We consider (a) weak $A = 0.5$, (b) intermediate $A = 1.0$, and (c) strong $A = 1.5$ SB coupling cases. Additional parameters for the isothermal and thermostatic processes are listed in Table~\ref{table:SimIsothermal}.
The potential surface with the eigenstates and eigenenergies is presented in Ref. \onlinecite{KT24JCP1}. In an isothermal process, the first excitation energy is $ \sim 0.8$, making the bath temperature low at $T=1/(k_{\rm B}\beta)  \approx 0.3$ and high at $T=1/(k_{\rm B}\beta)  \approx 5.0$.
We present results in the quasi-static case with each step having a time duration of $\tau = 1.0 \times 10^4$, while the non-equilibrium case was discussed in Ref. \onlinecite{KT24JCP2}.

In our previous papers,\cite{KT24JCP1,KT24JCP2} we introduced the quasi-static Massieu potential and dimensionless Clausius entropy defined as:
\begin{eqnarray}
\label{eq:Massieu}
\frac{d \Phi_{\rm A}^{\rm qst} ( t )}{d t} = - U_{\rm A}^{\rm qst} ( t ) \frac{d \beta^{\rm qst} ( t )}{d t}
- \tilde{E}^{\rm qst} ( t ) \frac{d P_{\rm A}^{\rm qst} ( t )}{d t} ,
\end{eqnarray}
and
\begin{eqnarray}
\label{eq:DLEntropy}
\frac{d \Lambda_{\rm A}^{\rm qst} ( t )}{d t} = \beta^{\rm qst} ( t ) \frac{d U_{\rm A}^{\rm qst} ( t )}{d t}
- \tilde{E}^{\rm qst} ( t ) \frac{d P_{\rm A}^{\rm qst} ( t )}{d t} ,
\end{eqnarray}
respectively, where $\tilde{E} ( t ) = \beta ( t ) E ( t )$, and the superscript ${\rm qst}$ indicates that the process is quasi-static. The dimensionless Clausius entropy is related to the quasi-static entropy by $S_{\rm A}^{\rm qst} ( t ) = k_{\rm B} \Lambda_{\rm A}^{\rm qst} ( t )$, allowing us to evaluate the quasi-static entropy using Eq. \eqref{eq:DLEntropy}.

To monitor the performance of the thermostatic Stirling engine, we introduced intensive work and heat\cite{KT24JCP1, KT24JCP2} as follows:
\begin{eqnarray}
\label{eq:Wint}
\frac{d W^{int}_{\rm A} ( t )}{d t} = \mathrm{tr}_{\rm A} \left\{ \frac{\partial \hat{H}_{\rm A} ( t )}{\partial t}
\hat{\rho}_{\rm A} ( t ) \right\} ,
\end{eqnarray}
and
\begin{eqnarray}
\label{eq:QA}
\frac{d Q_{\rm A} ( t )}{d t} = \mathrm{tr}_{\rm A}
\left\{ \hat{H}_{\rm A} ( t ) \frac{\partial \hat{\rho}_{\rm A} ( t )}{\partial t} \right\} .
\end{eqnarray}

We provide the demonstration program to evaluate these variables from Eqs. \eqref{eq:Enthalpy} and \eqref{eq:InternalEnergy} in the supporting information.

\renewcommand{\arraystretch}{1.3}

\begin{table}[!t]
\caption{\label{Cycelarameter}Time evolutions of the external force [$E ( t )$] and temperature [$T( t )$] in a four-step thermostatic Stirling engine with equal time intervals $\tau$. The cycle consists of (i) a hot isothermal process, (ii) a thermostatic transition from hot to cold, (iii) a cold isothermal process, and (iv) a thermostatic transition from cold to hot. We set $\Delta E = E_2 - E_1$ and $\Delta \beta = \beta_{\rm C} - \beta_{\rm H}$.}
\begin{ruledtabular}
\begin{tabular}{lcc}
&   $E ( t ) $ & $\beta ( t )$
\\
\hline
(i) hot isothermal & $E_1 + \Delta E t/\tau$ & $\beta_{\rm H}$
\\
(ii) hot to cold & $E_2$ & $\beta_{\rm H} + \Delta \beta (t/\tau - 1)$
\\
(iii) cold isothermal  & $E_2 - \Delta E (t/\tau - 2)$ & $\beta_{\rm C}$
\\
(iv) cold to hot  & $E_1$ & $\beta_{\rm C} - \Delta \beta (t/\tau - 3)$
\\
\end{tabular}
\end{ruledtabular}
\end{table}

\begin{figure}

\includegraphics[width=7cm]{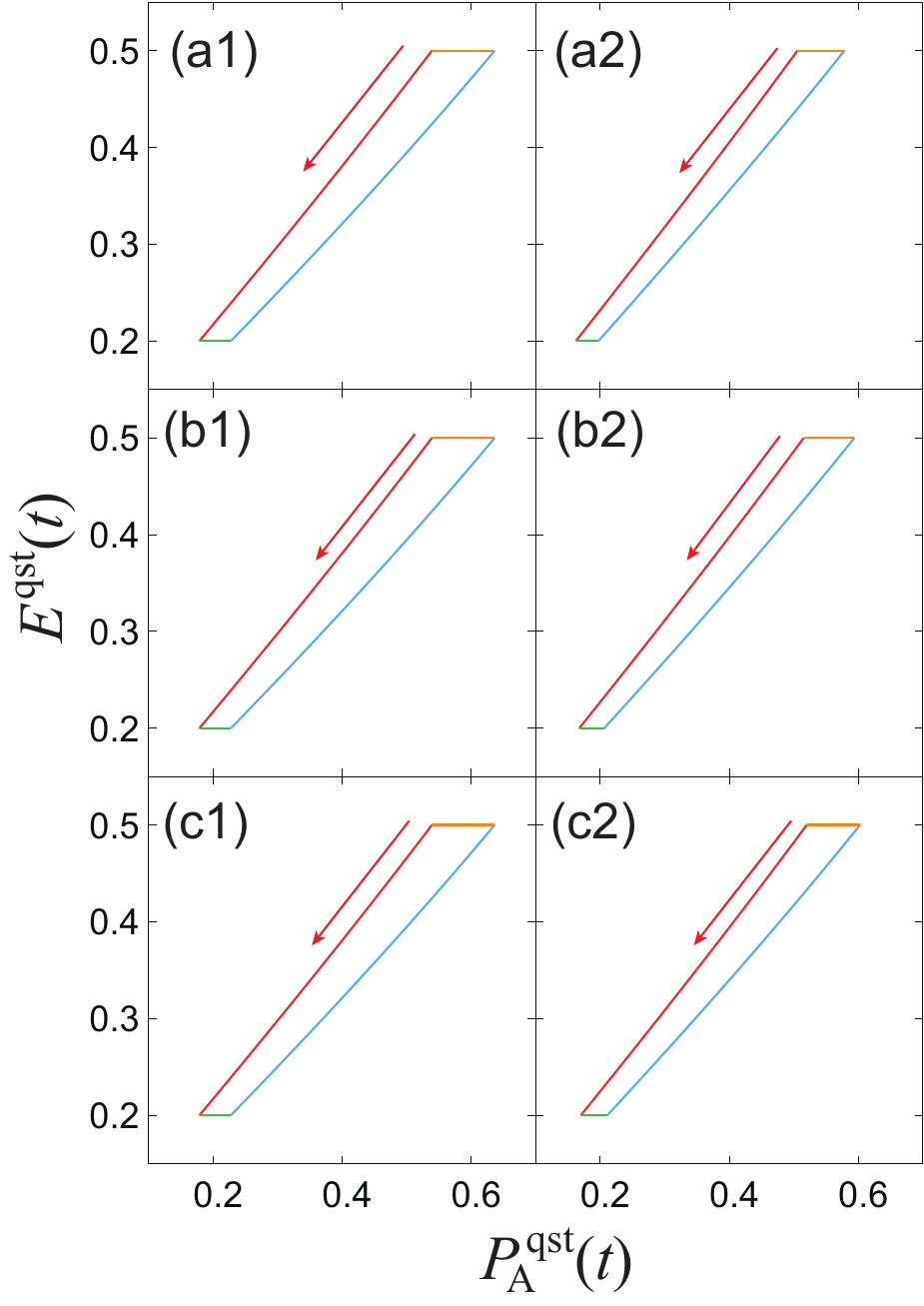}
\caption{\label{fig:EP} $E^{\rm qst} (t)$--$P_{\rm A}^{\rm qst}  ( t )$ diagrams for the thermostatic Stirling engine in the classical case (1, left column) and quantum case (2, right column) for (a) $A = 0.5$ (weak), (b) $1.0$ (intermediate), and (c) $1.5$ (strong) SB coupling strengths. In each plot, the four curves (or lines) represent (i) hot isothermal (red), (ii) from hot to cold thermostatic (green), (iii) cold isothermal (blue), and (iv) from cold to hot thermostatic (orange) processes, respectively.}
\end{figure}

\begin{figure}

\includegraphics[width=7cm]{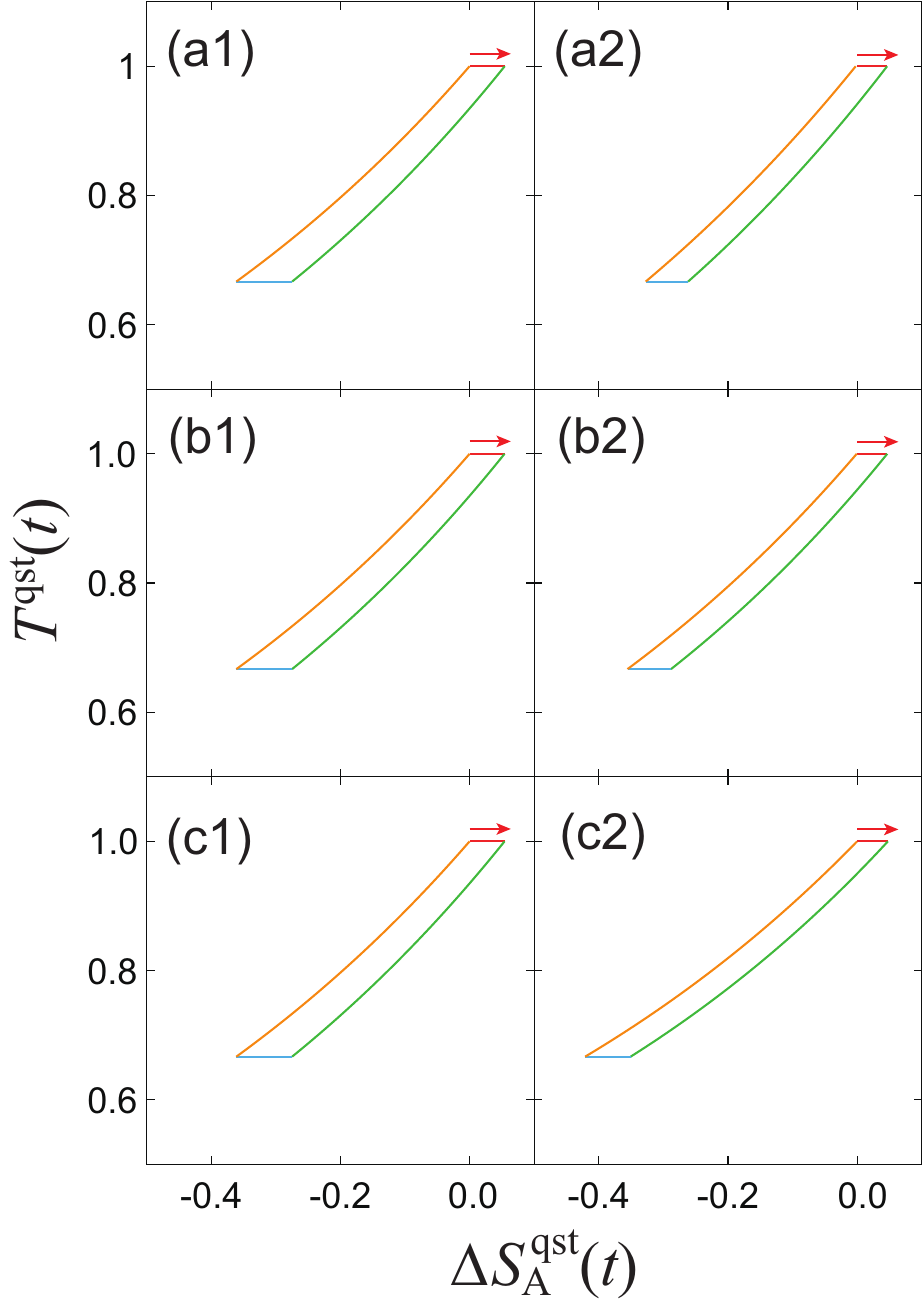}

\caption{\label{fig:TS} $T^{\rm qst} ( t )$--$S_{\rm A}^{\rm qst} ( t )$ diagrams for the thermostatic Stirling engine in the classical case (1, left column) and quantum case (2, right column) for (a) $A = 0.5$ (weak), (b) $1.0$ (intermediate), and (c) $1.5$ (strong) SB coupling strengths, respectively. Each cycle starts with the red arrow, and the four curves represent (i) hot isothermal (red), (ii) from hot to cold thermostatic (green), (iii) cold isothermal (blue), and (iv) from cold to hot thermostatic (orange) processes, respectively.}

\end{figure}

To elucidate the characteristics of a cyclic process, we constructed thermodynamic work diagrams for external forces, temperature, and their conjugate variables as the $E^{\rm qst}$--$P_{\rm A}^{\rm qst}$  and $T^{\rm qst}$--$S_{\rm A}^{\rm qst}$ diagrams, analogous to the $P$--$V$ diagram for a gas system. 
The work diagrams in the non-equilibrium state are given in Ref. \onlinecite{KT24JCP2}.
Figures \ref{fig:EP} and \ref{fig:TS} show the $E^{\rm qst}$--$P_{\rm A}^{\rm qst}$ and $T^{\rm qst}$--$S_{\rm A}^{\rm qst}$ diagrams for weak and strong SB coupling strengths in the classical (left column) and quantum (right column) cases. The trajectories of the work diagrams are periodic and closed because $P_{\rm A}^{\rm qst}$ and $S_{\rm A}^{\rm qst}$ are state variables. 

The processes in the $E^{\rm qst}$--$P_{\rm A}^{\rm qst}$ diagrams evolve counterclockwise, while those in the $T^{\rm qst}$--$S_{\rm A}^{\rm qst}$ diagrams evolve clockwise. Compared to the $P$--$V$ diagram for an ideal gas, the rotation directions in the $E^{\rm qst}$--$P_{\rm A}^{\rm qst}$ diagram are opposite because the signs of $P d V$ and $E^{\rm qst} d P_{\rm A}^{\rm qst}$ are opposite in the differential forms of the internal energy in ideal gas and dipole systems.
The area enclosed by each diagram corresponds to the positive work performed by the external field in clockwise evolution. We summarize the intensive work (Eq. \eqref{eq:Wint}) performed in one cycle for various SB couplings in the classical and quantum cases in Table \ref{table:CycleWork}.  For the $T^{\rm qst}$--$S_{\rm A}^{\rm qst}$ diagrams in Fig.~\ref{fig:TS}, the area enclosed by the clockwise curve corresponds to the heat, $Q_{\rm A}$ per cycle. The entropy changes monotonically in the thermostatic processes due to the concavity of the quasi-static Gibbs energy for $T$.\cite{KT24JCP1}

In the classical case, the equilibrium distribution is independent of SB coupling strength, as indicated by the Kramers equation. Therefore, the diagram is the same for the coupling strength. In the quantum case, due to bathentanglement, the trajectories change depending on SB coupling strength,\cite{KT22JCP1,KT22JCP2} but the difference is not significant because the temperature is not low, making the results closer to the classical case (see also Table \ref{table:CycleWork}).  

Compared to the non-equilibrium case,\cite{KT24JCP2} the trajectories in the quasi-static case are stable and do not fluctuate, even when the interaction is weak.

\begin{table}

\centering
\caption{\label{table:CycleWork} The intensive work performed in one cycle are shown for weak ($A = 0.5$), intermediate ($A = 1.0$), and strong ($A = 1.5$) SB coupling in both the classical and quantum cases.}

\begin{tabular}{ccc}
\hline
$\hspace{0.5cm} A \hspace{0.5cm}$ & classical & quantum
\\
\hline \hline
$0.5$ & \hspace{0.2cm}$-2.264 \times 10^{-2}$\hspace{0.2cm} & 
\hspace{0.2cm} $-1.657 \times 10^{-2}$ \hspace{0.2cm}
\\
$1.0$ & $-2.262 \times 10^{-2}$ & $-1.809 \times 10^{-2}$
\\
$1.5$ & $-2.258 \times 10^{-2}$ & $-1.913 \times 10^{-2}$
\\
\hline
\end{tabular}

\end{table}

In the quantum case, the work performed by the external field decreases with larger SB coupling strength due to bathentanglement, while in the classical case, it does not. Our previous study based on the spin-Boson model for the Carnot cycle\cite{KT22JCP2} showed that the intensive work performed in one cycle was independent of the SB coupling strength because the initial and final equilibrium states in isothermal processes were the same regardless of the SB coupling strength. Thus, from the Kelvin--Planck statement, the intensive work performed in the isothermal processes was independent of the SB coupling strength. However, in the present case, because the initial and final equilibrium states in isothermal processes differ depending on the SB coupling strength due to bathentanglement, as shown in Appendix \ref{sec:anhamonic}, the intensive work changes depending on the coupling strength.

Finally, in the classical and high-temperature limits, the T-QFPE results are equivalent to those obtained from the Langevin approach, where a Markovian description is applicable. However, at low temperatures, due to the bathentanglement, the subsystem follows non-factorial and non-Markovian dynamics, and its equilibrium state deviates from the Boltzmann distribution. This indicates that dynamics in a fully quantum regime cannot be described by the fluctuation theorem\cite{RevModPhys.81.1665,RevModPhys.83.771,sagawa20232law} and stochastic thermodynamics.\cite{Schmiedl_2008,Stochastic_Seifert_2012,Esposito2011,strasberg2022quantum,RaoEsposit2018,PhysRevE.105.064124} In other words, the difference between the classical and quantum results represents a breakdown of these theory in the fully quantum regime.

\begin{table}[!t]
\caption{\label{table:SimIsothermal} Parameter values used for the simulations of the Stirling engine. Here, $d x$ and $d p$ are the mesh sizes for position and momentum, respectively, in  Wigner space. The integers $N$ and $K$ are the cutoff numbers used in the T-QFPE.}
\begin{ruledtabular}
\begin{tabular}{lccccc}
& $A$  & $ N $ & $ K $ &  \hspace{3mm}$d x$\hspace{3mm}  &  \hspace{3mm}$d p$\hspace{3mm}
\\
\hline
\multirow{3}{*}{Classical} & $0.5$ & $\cdots$ & $\cdots$ & $0.25$ & $0.25$
\\
& $1.0$ & $\cdots$ & $\cdots$ & $0.25$ & $0.25$
\\
& $1.5$ & $\cdots$ & $\cdots$ & $0.25$ & $0.25$
\\
\hline
\multirow{3}{*}{Quantum} & $0.5$ & $6$ & $2$ & $0.3$ & $0.5$
\\
& $1.0$ & $7$ & $2$ & $0.3$ & $0.5$

\\
& $1.5$ & $8$ & $2$ & $0.3$ & $0.5$
\\
\end{tabular}
\end{ruledtabular}
\end{table}

\section{Conclusion}
\label{sec:conclude} 
 
This study presents a flexible and reliable simulation tool (T-QFPE) designed to enable non-experts in open quantum dynamics theory to quantitatively develop nonequilibrium thermodynamic theories. To ensure numerical reliability and facilitate the determination of working variables such as mesh size and time step, we included a non-Markovian test routine that compares the numerical results with analytical solutions for a harmonic Brownian system.

The central principle of thermodynamics is that thermal phenomena are described as intensive and extensive variables. Consequently, we introduce extensive variables conjugate to intensive variables as time-dependent physical observables.\cite{KT24JCP1,KT24JCP2}  
A routine for calculating the Helmholtz energy in the quasi-static case is included. As a demonstration, a thermostatic Stirling engine for an anharmonic Brownian system is simulated and analyzed in both quantum and classical cases. Through work diagrams, it was shown that due to bathentanglement, the work done on the system by the external field is smaller in the quantum case than in the classical case.

Since the thermodynamic Kramers equation used in this study is equivalent to the Langevin equation for thermostatic processes, it is also possible to explore more complex systems using molecular dynamics simulations in classical cases.

Although our current discussion focuses on an anharmonic Brownian system, the code can be extended to study chemical reaction systems,\cite{TW91PRA,TW92JCP,Shi2011PT,ZBT20JCP} ratchet systems,\cite{KT13JPCB} resonant tunneling systems,\cite{ST13JPSJ,ST14NJP,GST16JPSJ} non-adiabatic transition systems,\cite{IT17JCP,IDT19JCP,IT19JCTC} and vibrational modes of liquid water,\cite{IIT15JCP,IT16JCP,TT23JCP1,TT23JCP2}
which have been previously studied with the quantum hierarchical Fokker--Planck approach. This extension, for example, would allow for the investigation of the spatio-temporal distribution of entropy production in ratcheting systems.  

When focusing solely on quantum properties, simpler systems such as spin--boson systems are less computationally demanding and easier to analyze. A computer code for the thermodynamic spin--boson system will be presented in a forthcoming paper.\cite{KT24JCP4}

\section*{Supplementary Material}
Numerical integration codes for the T-QFPE and four demo codes (three for non-Markovian tests and one for thermostatic Stirling engine) are provided as supplemental materials. The manual can be found in the ReadMe.pdf file.

\section*{Acknowledgments}
Y.T. was supported by JSPS KAKENHI (Grant No.~B21H01884).
S.K. was supported by Grant-in-Aid for JSPS Fellows (Grant No.~24KJ1373).

\section*{Author declarations}
\subsection*{Conflict of Interest}
The authors have no conflicts to disclose.

\section*{Author Contributions}
{\bf Shoki Koyanagi}: Investigation (equal); Software (lead); Writing – original draft (equal). {\bf Yoshitaka Tanimura}: Conceptualization (lead); Investigation (equal); Writing – review and editing (lead).

\section*{Data availability}
The data that support the findings of this study are available from the corresponding author upon reasonable request.

\appendix

\section{The adiabatic transition and the translational symmetry}
\label{sec:AdditionalTerm}

When considering the adiabatic transition process, where the SB coupling strength $A ( t )$ depends on time, an additional term appears expressed as:
\begin{eqnarray}
\label{eq:AdditionalTerm}
A ( t ) \frac{d A ( t )}{d t} q \frac{\partial}{\partial p} W_{\vec{n}} ( p , q ; t ) ,
\end{eqnarray}
on the right side of Eq. \eqref{eq:T-QFPE}. This term violates the translational symmetry of the reduced system because Eq. \eqref{eq:AdditionalTerm} is proportional to the position $q$, and $d A ( t ) / d t$ is not negligible unless $A(t)$ changes very slowly.  This is due to the ambiguity in defining the noise correlation time of a heat bath when $A(t)$ is time-dependent. However, in our current study, $A(t)$ is constant, so this term vanishes.

\section{Numerical implementation of the T-QFPE}
\label{sec:Detail} 

As with the LT-QFPE,\cite{IT19JCTC} the T-QFPE are simultaneous differential equations expressed in terms of the reduced density matrix elements or WDF.
Due to the complex hierarchical structure, especially at low temperatures, the numerical integration of the T-QFPE is computationally intensive in both memory and central processing unit (CPU) efficiency. Efforts have been made to reduce computational costs by improving the algorithmic and numerical techniques.\cite{T20JCP} For example, continuous efforts have been made to construct efficient hierarchy elements.\cite{10.1063/5.0007327,PhysRevLett.129.230601,10.1063/5.0095961,10.1063/5.0209348,10.1063/5.0214051} Here, we use the Pad\'e spectral decompositions\cite{YanPade10A,YanPade10B,YanPade11} instead of the Matsubara frequency decomposition.\cite{T90PRA,IT05JPSJ,T14JCP,T15JCP} In our source code, we employed the Pad\'e factors listed in Ref. \onlinecite{IT19JCTC}.

Although not used here, if we wish to handle larger systems, such as a system with conical intersection,\cite{IT18CP}
a numerical algorithm based on the optimization of hierarchical basis sets\cite{YanOptimalBasis2015} and a tensor network algorithm to reduce the number of calculations can be employed.\cite{ShiJCP2018Tensor,BorrelliCP2018Tensor}

\subsection{Integration Routine}

We integrate Eq. \eqref{eq:T-QFPE} using the predictor--corrector approach in conjunction with the Adams--Bashforth and Adams--Moulton methods.\cite{press1988numerical}  
In this approach, we first compute an initial guess of the auxiliary WDFs at the next time step $t + \delta t$, as the ``predictor'' using the fourth-order Adams--Bashforth method expressed as follows:
\begin{eqnarray}
\label{eq:AB4}
&& W^{\rm pred}_{\vec{n}} ( t + \delta t ) = W_{\vec{n}} ( t ) + \frac{\delta t}{24} \left[ 55 K_{\vec{n}} ( t ) - 59 K_{\vec{n}} ( t - \delta t ) \right. \nonumber \\
&& \qquad 
\left. + 37 K_{\vec{n}} ( t - 2 \delta t ) - 9 K_{\vec{n}} ( t - 3 \delta t ) \right],
\end{eqnarray}
where $K_{\vec{n}} ( t ) = \hat{\mathcal{L}}_{\vec{n}} ( t ) W_{\vec{n}} ( p , q ; t )$ describes the time evolution of each $W_{\vec{n}} ( p , q ; t )$ using the Liouvillian for the ${\vec{n}}$ element expressed as $\hat{\mathcal{L}}_{\vec{n}} ( t )$.  Then, 
to refine the initial guess, we compute the $W_{\vec{n}} ( p , q ; t + \delta t )$, as the ``corrector'' using the fourth-order Adams--Moulton method expressed as:
\begin{eqnarray}
\label{eq:AM4}
&& W_{\vec{n}} ( p , q ; t + \delta t )  = W_{\vec{n}} ( p , q ; t )  \nonumber \\
&& \qquad + \frac{\delta t}{24} 
\left[ 9 K^{\rm pred}_{\vec{n}} ( t + \delta t ) + 19 K_{\vec{n}} ( t ) \right. \nonumber \\
&&
\qquad \qquad \left.
\qquad - 5 K_{\vec{n}} ( t - \delta t ) + K_{\vec{n}} ( t - 2 \delta t ) \right],
\end{eqnarray}
where $K_{\vec{n}}^{\rm pred} ( t ) = \hat{\mathcal{L}}_{\vec{n}} ( t ) W^{\rm pred}_{\vec{n}} ( p , q ; t )$. In the predictor--corrector method, the corrector $W_{\vec{n}} ( p , q ; t + \delta t )$ is regarded as the auxiliary WDF at time $t + \delta t$.

To adapt the predictor--corrector approach, the initial values $W_{\vec{n}} ( t-\delta t)$, $W_{\vec{n}} ( t-2\delta t)$, and $W_{\vec{n}} ( t-3\delta t)$ must be prepared. Thus, we use the fourth-order Runge--Kutta method for the first three steps from $t-3\delta t$ to evaluate the initial values. 

The Kramers Eq. \eqref{heom_cl} with Eq. \eqref{L_cl} is also integrated accordingly.

\subsection{Open Multi--Processing (OMP) and Compute Unified Device Architecture (CUDA)}

Because the T-QFPE [Eq. \eqref{eq:T-QFPE}] and the Kramers equation [Eq. \eqref{heom_cl}] are linear differential equations, parallelizing the routines enhances computational performance. Here, we use two parallelization technologies: Open Multi--Processing (OMP), which parallelizes the CPU cores,\cite{doi:10.1021/ct3003833,doi:10.1021/ct500629s,KRAMER2018262} and Compute Unified Device Architecture (CUDA), which performs parallelization for the Graphic Processing Unit (GPU). \cite{doi:10.1021/ct200126d,Hein_2012,TT15JCTC,IT19JCTC}

In the supporting information, we provide two types of T-QFPE programs: one using only OMP and one using both OMP and CUDA technologies. To use CUDA, remove the comment symbol ``//'' from the third line ``//\#define UseCUDA'' in the ``MAIN.h'' file.

\section{Partition function of anharmonic potential}
\label{sec:anhamonic}

We consider an anharmonic subsystem expressed as:
\begin{eqnarray}
U ( q ) = \frac{1}{2} m \omega_0^2 q^2 - E q + U'( q ) ,
\end{eqnarray}
where $U'( q )$ is the anharmonic part of the potential. The partition function of any anharmonic Brownian oscillator system can be evaluated using the imaginary HEOM approaches.\cite{T15JCP}
Alternatively, when the anharmonicity is weak, the partition function can also be evaluated using the generating functional approach  as:\cite{OT96JCP,TO97JCP,T15JCP} 
\begin{eqnarray}
\label{eq:GeneratingFunctional1}
\begin{split}
&Z_{\rm anh}^{\rm eq} = \left[ 1 -  \frac1{\hbar} \int_0^{\beta \hbar}d\tau' U'\left(\hbar \frac{\partial}{\partial \bar J(\tau')} \right) \right. \\ 
& +
 \frac1{2\hbar^2} \int_0^{\beta \hbar} \int_0^{\beta \hbar} d\tau'' d\tau' U'\left(\hbar \frac{\partial}{ \partial \bar J(\tau')} \right)
 U'\left( \hbar \frac{\partial}{\partial \bar J(\tau'')} \right) \\ 
 &
  +  \ldots \left]
 Z_0 [\bar{J} ; \beta \hbar ]  \right|_{\bar J (\tau) = E} \nonumber,
\end{split}
\\
\end{eqnarray}
where $Z_0 [ \bar{J} ; \beta \hbar ]$ is the generating functional of the subsystem defined by the potential
\begin{eqnarray}
U ( q , \tau ) = \frac{1}{2} m \omega_0^2 q^2 - \bar{J} ( \tau ) q .
\end{eqnarray}
Because $U ( q , \tau ) $ is harmonic, we can obtain the generating functional expressed as\cite{OT96JCP} 
\begin{eqnarray}
\label{eq:GeneratingFunctional2}
\begin{split}
&Z_0 [ \bar{J} ; \beta \hbar ] 
= \frac{1}{\beta \hbar \omega_0} \prod_{n = 1}^\infty \frac{\nu_n^2}{\omega_0^2 + \nu_n + \zeta_n} \\
&
\times
\exp \left[ \frac{1}{2 \hbar M} \int^{\beta \hbar}_0 d \tau \int^{\beta \hbar}_0 d \sigma \bar{J} ( \tau )
\bar{J} ( \sigma ) \Lambda ( \tau - \sigma ) \right] ,
\end{split}
\end{eqnarray}
where 
\begin{eqnarray}
\label{eq:GeneratingFunctional3}
\Lambda ( \tau ) = \frac{1}{\beta \hbar} \sum_{n = - \infty}^\infty
\frac{e^{i \nu_n \tau}}{\omega_0^2 + \nu_n^2 + \zeta_n} ,
\end{eqnarray}
and
\begin{eqnarray}
\label{eq:GeneratingFunctional4}
\zeta_n = \frac{1}{m} \int^\infty_0 \frac{J ( \omega )}{\hbar \omega} \frac{2 \nu_n^2}{ \omega^2 + \nu_n^2}
d \omega .
\end{eqnarray}
As indicated in the above expression, the value of the partition function in the quantum case changes as a function of the SB coupling strength through $J(\omega)$, even in the harmonic case, while in the classical case, it remains unchanged.

To apply the above expression, both the anharmonicity and the external field are assumed to be weak.
The anharmonic part of the potential is given by:
\begin{eqnarray}
\label{eq:AnharmonicPotential}
U' ( q ) = U_3 q^3 + U_4 q^4.
\end{eqnarray}
From Eqs. \eqref{eq:GeneratingFunctional1}-\eqref{eq:GeneratingFunctional4}, we can evaluate the intensive work done in one cycle of the Stirling engines. The difference between classical and quantum values in the Ohmic case is evaluated as:
\begin{eqnarray}
\label{eq:WorkDiff}
\begin{split}
& W_{\rm QM}^{int} - W_{\rm CL}^{int} \\
& \qquad = \frac{3 \hbar U_3}{m} ( \beta_{\rm H} \Lambda_{\beta_{\rm H}} ( 0 )- \beta_{\rm C} \Lambda_{\beta_{\rm C}} ( 0 ) )
( \epsilon_2 - \epsilon_1 ) \\
& \qquad 
+ \frac{6 \hbar U_4}{m} ( \beta_{\rm H} \Lambda_{\beta_{\rm H}} ( 0 )- \beta_{\rm C} \Lambda_{\beta_{\rm C}} ( 0 ))
( \epsilon_2^2 - \epsilon_1^2 ) ,
\end{split}
\end{eqnarray}
where $\Lambda_{\beta_{\rm H}} ( 0 )$ and $\Lambda_{\beta_{\rm C}} ( 0 )$ are $\Lambda( 0 )$ at $\beta_{\rm H}$ and $\beta_{\rm C}$, and $\epsilon_\alpha = E_\alpha / m \omega_0^2 \; ( \alpha = 1 , 2 )$. For $U_3 , U_4 > 0$, the right--hand side of Eq. \eqref{eq:WorkDiff} is positive for $E_1 > E_2$ because $\beta_{\rm H} \Lambda_{\beta_{\rm H}} (0) - \beta_{\rm C} \Lambda_{\beta_{\rm C}} (0)$ becomes negative.  This implies $W_{\rm QM}^{int} > W_{\rm CL}^{int}$. Since $\Lambda$ becomes small with increasing the SB coupling $A$, the quantum result approaches the classical result for $A \rightarrow \infty$.

Although our numerical results presented in Sec. \ref{sec:QSE} are not in the perturbative regime, the results in Table \ref{table:CycleWork} 
can be qualitatively explained by this argument.  

\bibliography{references,tanimura_publist}

\begin{thebibliography}{88}%
\makeatletter
\providecommand \@ifxundefined [1]{%
 \@ifx{#1\undefined}
}%
\providecommand \@ifnum [1]{%
 \ifnum #1\expandafter \@firstoftwo
 \else \expandafter \@secondoftwo
 \fi
}%
\providecommand \@ifx [1]{%
 \ifx #1\expandafter \@firstoftwo
 \else \expandafter \@secondoftwo
 \fi
}%
\providecommand \natexlab [1]{#1}%
\providecommand \enquote  [1]{``#1''}%
\providecommand \bibnamefont  [1]{#1}%
\providecommand \bibfnamefont [1]{#1}%
\providecommand \citenamefont [1]{#1}%
\providecommand \href@noop [0]{\@secondoftwo}%
\providecommand \href [0]{\begingroup \@sanitize@url \@href}%
\providecommand \@href[1]{\@@startlink{#1}\@@href}%
\providecommand \@@href[1]{\endgroup#1\@@endlink}%
\providecommand \@sanitize@url [0]{\catcode `\\12\catcode `\$12\catcode
  `\&12\catcode `\#12\catcode `\^12\catcode `\_12\catcode `\%12\relax}%
\providecommand \@@startlink[1]{}%
\providecommand \@@endlink[0]{}%
\providecommand \url  [0]{\begingroup\@sanitize@url \@url }%
\providecommand \@url [1]{\endgroup\@href {#1}{\urlprefix }}%
\providecommand \urlprefix  [0]{URL }%
\providecommand \Eprint [0]{\href }%
\providecommand \doibase [0]{https://doi.org/}%
\providecommand \selectlanguage [0]{\@gobble}%
\providecommand \bibinfo  [0]{\@secondoftwo}%
\providecommand \bibfield  [0]{\@secondoftwo}%
\providecommand \translation [1]{[#1]}%
\providecommand \BibitemOpen [0]{}%
\providecommand \bibitemStop [0]{}%
\providecommand \bibitemNoStop [0]{.\EOS\space}%
\providecommand \EOS [0]{\spacefactor3000\relax}%
\providecommand \BibitemShut  [1]{\csname bibitem#1\endcsname}%
\let\auto@bib@innerbib\@empty
\bibitem [{\citenamefont {Grabert}, \citenamefont {Schramm},\ and\
  \citenamefont {Ingold}(1988)}]{GRABERT1988115}%
  \BibitemOpen
  \bibfield  {author} {\bibinfo {author} {\bibfnamefont {H.}~\bibnamefont
  {Grabert}}, \bibinfo {author} {\bibfnamefont {P.}~\bibnamefont {Schramm}},\
  and\ \bibinfo {author} {\bibfnamefont {G.-L.}\ \bibnamefont {Ingold}},\
  }\bibfield  {title} {\enquote {\bibinfo {title} {Quantum brownian motion: The
  functional integral approach},}\ }\href
  {https://doi.org/https://doi.org/10.1016/0370-1573(88)90023-3} {\bibfield
  {journal} {\bibinfo  {journal} {Physics Reports}\ }\textbf {\bibinfo {volume}
  {168}},\ \bibinfo {pages} {115--207} (\bibinfo {year} {1988})}\BibitemShut
  {NoStop}%
\bibitem [{\citenamefont {Weiss}(2012)}]{Weiss2012}%
  \BibitemOpen
  \bibfield  {author} {\bibinfo {author} {\bibfnamefont {U.}~\bibnamefont
  {Weiss}},\ }\href {https://doi.org/10.1142/8334} {\emph {\bibinfo {title}
  {Quantum Dissipative Systems}}},\ \bibinfo {edition} {4th}\ ed.\ (\bibinfo
  {publisher} {WORLD SCIENTIFIC},\ \bibinfo {year} {2012})\BibitemShut
  {NoStop}%
\bibitem [{\citenamefont {Tanimura}(2006)}]{T06JPSJ}%
  \BibitemOpen
  \bibfield  {author} {\bibinfo {author} {\bibfnamefont {Y.}~\bibnamefont
  {Tanimura}},\ }\bibfield  {title} {\enquote {\bibinfo {title} {Stochastic
  \uppercase{L}iouville, \uppercase{L}angevin,
  \uppercase{F}okker-\uppercase{P}lanck, and master equation qpproaches to
  quantum dissipative systems},}\ }\href
  {https://doi.org/10.1143/JPSJ.75.082001} {\bibfield  {journal} {\bibinfo
  {journal} {Journal of the Physical Society of Japan}\ }\textbf {\bibinfo
  {volume} {75}},\ \bibinfo {pages} {082001} (\bibinfo {year}
  {2006})}\BibitemShut {NoStop}%
\bibitem [{\citenamefont {Tanimura}(2020)}]{T20JCP}%
  \BibitemOpen
  \bibfield  {author} {\bibinfo {author} {\bibfnamefont {Y.}~\bibnamefont
  {Tanimura}},\ }\bibfield  {title} {\enquote {\bibinfo {title} {Numerically
  "exact" approach to open quantum dynamics: The hierarchical equations of
  motion (\uppercase{HEOM})},}\ }\href {https://doi.org/10.1063/5.0011599}
  {\bibfield  {journal} {\bibinfo  {journal} {The Journal of Chemical Physics}\
  }\textbf {\bibinfo {volume} {153}},\ \bibinfo {pages} {020901} (\bibinfo
  {year} {2020})}\BibitemShut {NoStop}%
\bibitem [{\citenamefont {Tanimura}\ and\ \citenamefont
  {Kubo}(1989)}]{TK89JPSJ1}%
  \BibitemOpen
  \bibfield  {author} {\bibinfo {author} {\bibfnamefont {Y.}~\bibnamefont
  {Tanimura}}\ and\ \bibinfo {author} {\bibfnamefont {R.}~\bibnamefont
  {Kubo}},\ }\bibfield  {title} {\enquote {\bibinfo {title} {Time evolution of
  a quantum system in contact with a nearly
  \uppercase{G}aussian-\uppercase{M}arkoffian noise bath},}\ }\href
  {https://doi.org/10.1143/JPSJ.58.101} {\bibfield  {journal} {\bibinfo
  {journal} {Journal of the Physical Society of Japan}\ }\textbf {\bibinfo
  {volume} {58}},\ \bibinfo {pages} {101--114} (\bibinfo {year}
  {1989})}\BibitemShut {NoStop}%
\bibitem [{\citenamefont {Tanimura}(1990)}]{T90PRA}%
  \BibitemOpen
  \bibfield  {author} {\bibinfo {author} {\bibfnamefont {Y.}~\bibnamefont
  {Tanimura}},\ }\bibfield  {title} {\enquote {\bibinfo {title}
  {Nonperturbative expansion method for a quantum system coupled to a
  harmonic-oscillator bath},}\ }\href
  {https://doi.org/10.1103/PhysRevA.41.6676} {\bibfield  {journal} {\bibinfo
  {journal} {Phys. Rev. A}\ }\textbf {\bibinfo {volume} {41}},\ \bibinfo
  {pages} {6676--6687} (\bibinfo {year} {1990})}\BibitemShut {NoStop}%
\bibitem [{\citenamefont {Ishizaki}\ and\ \citenamefont
  {Tanimura}(2005)}]{IT05JPSJ}%
  \BibitemOpen
  \bibfield  {author} {\bibinfo {author} {\bibfnamefont {A.}~\bibnamefont
  {Ishizaki}}\ and\ \bibinfo {author} {\bibfnamefont {Y.}~\bibnamefont
  {Tanimura}},\ }\bibfield  {title} {\enquote {\bibinfo {title} {Quantum
  dynamics of system strongly coupled to low-temperature colored noise bath:
  \uppercase{R}educed hierarchy equations approach},}\ }\href
  {https://doi.org/10.1143/JPSJ.74.3131} {\bibfield  {journal} {\bibinfo
  {journal} {Journal of the Physical Society of Japan}\ }\textbf {\bibinfo
  {volume} {74}},\ \bibinfo {pages} {3131--3134} (\bibinfo {year}
  {2005})}\BibitemShut {NoStop}%
\bibitem [{\citenamefont {Tanimura}(2014)}]{T14JCP}%
  \BibitemOpen
  \bibfield  {author} {\bibinfo {author} {\bibfnamefont {Y.}~\bibnamefont
  {Tanimura}},\ }\bibfield  {title} {\enquote {\bibinfo {title} {Reduced
  hierarchical equations of motion in real and imaginary time: Correlated
  initial states and thermodynamic quantities},}\ }\href
  {https://doi.org/10.1063/1.4890441} {\bibfield  {journal} {\bibinfo
  {journal} {The Journal of Chemical Physics}\ }\textbf {\bibinfo {volume}
  {141}},\ \bibinfo {pages} {044114} (\bibinfo {year} {2014})}\BibitemShut
  {NoStop}%
\bibitem [{\citenamefont {Tanimura}(2015)}]{T15JCP}%
  \BibitemOpen
  \bibfield  {author} {\bibinfo {author} {\bibfnamefont {Y.}~\bibnamefont
  {Tanimura}},\ }\bibfield  {title} {\enquote {\bibinfo {title} {Real-time and
  imaginary-time quantum hierarchal \uppercase{F}okker-\uppercase{P}lanck
  equations},}\ }\href {https://doi.org/10.1063/1.4916647} {\bibfield
  {journal} {\bibinfo  {journal} {The Journal of Chemical Physics}\ }\textbf
  {\bibinfo {volume} {142}},\ \bibinfo {pages} {144110} (\bibinfo {year}
  {2015})}\BibitemShut {NoStop}%
\bibitem [{\citenamefont {Tanimura}\ and\ \citenamefont
  {Wolynes}(1991)}]{TW91PRA}%
  \BibitemOpen
  \bibfield  {author} {\bibinfo {author} {\bibfnamefont {Y.}~\bibnamefont
  {Tanimura}}\ and\ \bibinfo {author} {\bibfnamefont {P.~G.}\ \bibnamefont
  {Wolynes}},\ }\bibfield  {title} {\enquote {\bibinfo {title} {Quantum and
  classical \uppercase{F}okker-\uppercase{P}lanck equations for a
  \uppercase{G}aussian-\uppercase{M}arkovian noise bath},}\ }\href
  {https://doi.org/10.1103/PhysRevA.43.4131} {\bibfield  {journal} {\bibinfo
  {journal} {Phys. Rev. A}\ }\textbf {\bibinfo {volume} {43}},\ \bibinfo
  {pages} {4131--4142} (\bibinfo {year} {1991})}\BibitemShut {NoStop}%
\bibitem [{\citenamefont {Tanimura}\ and\ \citenamefont
  {Wolynes}(1992)}]{TW92JCP}%
  \BibitemOpen
  \bibfield  {author} {\bibinfo {author} {\bibfnamefont {Y.}~\bibnamefont
  {Tanimura}}\ and\ \bibinfo {author} {\bibfnamefont {P.~G.}\ \bibnamefont
  {Wolynes}},\ }\bibfield  {title} {\enquote {\bibinfo {title} {The interplay
  of tunneling, resonance, and dissipation in quantum barrier crossing: A
  numerical study},}\ }\href {https://doi.org/10.1063/1.462301} {\bibfield
  {journal} {\bibinfo  {journal} {The Journal of Chemical Physics}\ }\textbf
  {\bibinfo {volume} {96}},\ \bibinfo {pages} {8485--8496} (\bibinfo {year}
  {1992})}\BibitemShut {NoStop}%
\bibitem [{\citenamefont {Kato}\ and\ \citenamefont
  {Tanimura}(2013)}]{KT13JPCB}%
  \BibitemOpen
  \bibfield  {author} {\bibinfo {author} {\bibfnamefont {A.}~\bibnamefont
  {Kato}}\ and\ \bibinfo {author} {\bibfnamefont {Y.}~\bibnamefont
  {Tanimura}},\ }\bibfield  {title} {\enquote {\bibinfo {title} {Quantum
  suppression of ratchet rectification in a \uppercase{B}rownian system driven
  by a biharmonic force},}\ }\href {https://doi.org/10.1021/jp403056h}
  {\bibfield  {journal} {\bibinfo  {journal} {The Journal of Physical Chemistry
  B}\ }\textbf {\bibinfo {volume} {117}},\ \bibinfo {pages} {13132--13144}
  (\bibinfo {year} {2013})},\ \Eprint {https://arxiv.org/abs/1305.1402}
  {arXiv:1305.1402} \BibitemShut {NoStop}%
\bibitem [{\citenamefont {Sakurai}\ and\ \citenamefont
  {Tanimura}(2011)}]{ST11JPCA}%
  \BibitemOpen
  \bibfield  {author} {\bibinfo {author} {\bibfnamefont {A.}~\bibnamefont
  {Sakurai}}\ and\ \bibinfo {author} {\bibfnamefont {Y.}~\bibnamefont
  {Tanimura}},\ }\bibfield  {title} {\enquote {\bibinfo {title} {Does $\hbar$
  play a role in multidimensional spectroscopy? reduced hierarchy equations of
  motion approach to molecular vibrations},}\ }\href
  {https://doi.org/10.1021/jp1095618} {\bibfield  {journal} {\bibinfo
  {journal} {The Journal of Physical Chemistry A}\ }\textbf {\bibinfo {volume}
  {115}},\ \bibinfo {pages} {4009--4022} (\bibinfo {year} {2011})}\BibitemShut
  {NoStop}%
\bibitem [{\citenamefont {Ikeda}\ and\ \citenamefont
  {Tanimura}(2019)}]{IT19JCTC}%
  \BibitemOpen
  \bibfield  {author} {\bibinfo {author} {\bibfnamefont {T.}~\bibnamefont
  {Ikeda}}\ and\ \bibinfo {author} {\bibfnamefont {Y.}~\bibnamefont
  {Tanimura}},\ }\bibfield  {title} {\enquote {\bibinfo {title}
  {Low-temperature quantum \uppercase{F}okker-\uppercase{P}lanck and
  \uppercase{S}moluchowski equations and their extension to multistate
  systems},}\ }\href {https://doi.org/10.1021/acs.jctc.8b01195} {\bibfield
  {journal} {\bibinfo  {journal} {Journal of Chemical Theory and Computation}\
  }\textbf {\bibinfo {volume} {15}},\ \bibinfo {pages} {2517--2534} (\bibinfo
  {year} {2019})}\BibitemShut {NoStop}%
\bibitem [{\citenamefont {Song}\ and\ \citenamefont
  {Shi}(2017)}]{PhysRevB.95.064308}%
  \BibitemOpen
  \bibfield  {author} {\bibinfo {author} {\bibfnamefont {L.}~\bibnamefont
  {Song}}\ and\ \bibinfo {author} {\bibfnamefont {Q.}~\bibnamefont {Shi}},\
  }\bibfield  {title} {\enquote {\bibinfo {title} {Hierarchical equations of
  motion method applied to nonequilibrium heat transport in model molecular
  junctions: Transient heat current and high-order moments of the current
  operator},}\ }\href {https://doi.org/10.1103/PhysRevB.95.064308} {\bibfield
  {journal} {\bibinfo  {journal} {Phys. Rev. B}\ }\textbf {\bibinfo {volume}
  {95}},\ \bibinfo {pages} {064308} (\bibinfo {year} {2017})}\BibitemShut
  {NoStop}%
\bibitem [{\citenamefont {Gong}\ \emph {et~al.}(2022)\citenamefont {Gong},
  \citenamefont {Wang}, \citenamefont {Zheng}, \citenamefont {Xu},\ and\
  \citenamefont {Yan}}]{YiJing2022}%
  \BibitemOpen
  \bibfield  {author} {\bibinfo {author} {\bibfnamefont {H.}~\bibnamefont
  {Gong}}, \bibinfo {author} {\bibfnamefont {Y.}~\bibnamefont {Wang}}, \bibinfo
  {author} {\bibfnamefont {X.}~\bibnamefont {Zheng}}, \bibinfo {author}
  {\bibfnamefont {R.}~\bibnamefont {Xu}},\ and\ \bibinfo {author}
  {\bibfnamefont {Y.}~\bibnamefont {Yan}},\ }\bibfield  {title} {\enquote
  {\bibinfo {title} {{Nonequilibrium work distributions in quantum impurity
  system–bath mixing processes}},}\ }\href
  {https://doi.org/10.1063/5.0095549} {\bibfield  {journal} {\bibinfo
  {journal} {The Journal of Chemical Physics}\ }\textbf {\bibinfo {volume}
  {157}},\ \bibinfo {pages} {054109} (\bibinfo {year} {2022})}\BibitemShut
  {NoStop}%
\bibitem [{\citenamefont {Latune}, \citenamefont {Pleasance},\ and\
  \citenamefont {Petruccione}(2023)}]{Petruccione2023}%
  \BibitemOpen
  \bibfield  {author} {\bibinfo {author} {\bibfnamefont {C.~L.}\ \bibnamefont
  {Latune}}, \bibinfo {author} {\bibfnamefont {G.}~\bibnamefont {Pleasance}},\
  and\ \bibinfo {author} {\bibfnamefont {F.}~\bibnamefont {Petruccione}},\
  }\bibfield  {title} {\enquote {\bibinfo {title} {Cyclic quantum engines
  enhanced by strong bath coupling},}\ }\href
  {https://doi.org/10.1103/PhysRevApplied.20.024038} {\bibfield  {journal}
  {\bibinfo  {journal} {Phys. Rev. Appl.}\ }\textbf {\bibinfo {volume} {20}},\
  \bibinfo {pages} {024038} (\bibinfo {year} {2023})}\BibitemShut {NoStop}%
\bibitem [{\citenamefont {Boettcher}\ \emph {et~al.}(2024)\citenamefont
  {Boettcher}, \citenamefont {Hartmann}, \citenamefont {Beyer},\ and\
  \citenamefont {Strunz}}]{Strunz2024}%
  \BibitemOpen
  \bibfield  {author} {\bibinfo {author} {\bibfnamefont {V.}~\bibnamefont
  {Boettcher}}, \bibinfo {author} {\bibfnamefont {R.}~\bibnamefont {Hartmann}},
  \bibinfo {author} {\bibfnamefont {K.}~\bibnamefont {Beyer}},\ and\ \bibinfo
  {author} {\bibfnamefont {W.~T.}\ \bibnamefont {Strunz}},\ }\bibfield  {title}
  {\enquote {\bibinfo {title} {{Dynamics of a strongly coupled quantum heat
  engine—Computing bath observables from the hierarchy of pure states}},}\
  }\href {https://doi.org/10.1063/5.0192075} {\bibfield  {journal} {\bibinfo
  {journal} {The Journal of Chemical Physics}\ }\textbf {\bibinfo {volume}
  {160}},\ \bibinfo {pages} {094108} (\bibinfo {year} {2024})}\BibitemShut
  {NoStop}%
\bibitem [{\citenamefont {Gatto}\ \emph {et~al.}(2024)\citenamefont {Gatto},
  \citenamefont {Colla}, \citenamefont {Heinz-Peter},\ and\ \citenamefont
  {Thoss}}]{Thoss2024}%
  \BibitemOpen
  \bibfield  {author} {\bibinfo {author} {\bibfnamefont {S.}~\bibnamefont
  {Gatto}}, \bibinfo {author} {\bibfnamefont {A.}~\bibnamefont {Colla}},
  \bibinfo {author} {\bibfnamefont {B.}~\bibnamefont {Heinz-Peter}},\ and\
  \bibinfo {author} {\bibfnamefont {M.}~\bibnamefont {Thoss}},\ }\bibfield
  {title} {\enquote {\bibinfo {title} {Quantum thermodynamics of the spin-boson
  model using the principle of minimal dissipation},}\ }\href
  {https://arxiv.org/abs/2404.12118} {\bibfield  {journal} {\bibinfo  {journal}
  {Phys. Rev. A}\ } (\bibinfo {year} {2024})}\BibitemShut {NoStop}%
\bibitem [{\citenamefont {Makri}(1995)}]{Makri95}%
  \BibitemOpen
  \bibfield  {author} {\bibinfo {author} {\bibfnamefont {N.}~\bibnamefont
  {Makri}},\ }\bibfield  {title} {\enquote {\bibinfo {title} {{Numerical path
  integral techniques for long time dynamics of quantum dissipative
  systems}},}\ }\href {https://doi.org/10.1063/1.531046} {\bibfield  {journal}
  {\bibinfo  {journal} {Journal of Mathematical Physics}\ }\textbf {\bibinfo
  {volume} {36}},\ \bibinfo {pages} {2430--2457} (\bibinfo {year}
  {1995})}\BibitemShut {NoStop}%
\bibitem [{\citenamefont {Makri}\ and\ \citenamefont
  {Makarov}(1995{\natexlab{a}})}]{Makri96}%
  \BibitemOpen
  \bibfield  {author} {\bibinfo {author} {\bibfnamefont {N.}~\bibnamefont
  {Makri}}\ and\ \bibinfo {author} {\bibfnamefont {D.~E.}\ \bibnamefont
  {Makarov}},\ }\bibfield  {title} {\enquote {\bibinfo {title} {{Tensor
  propagator for iterative quantum time evolution of reduced density matrices.
  I. Theory}},}\ }\href {https://doi.org/10.1063/1.469508} {\bibfield
  {journal} {\bibinfo  {journal} {The Journal of Chemical Physics}\ }\textbf
  {\bibinfo {volume} {102}},\ \bibinfo {pages} {4600--4610} (\bibinfo {year}
  {1995}{\natexlab{a}})}\BibitemShut {NoStop}%
\bibitem [{\citenamefont {Makri}\ and\ \citenamefont
  {Makarov}(1995{\natexlab{b}})}]{Makri96B}%
  \BibitemOpen
  \bibfield  {author} {\bibinfo {author} {\bibfnamefont {N.}~\bibnamefont
  {Makri}}\ and\ \bibinfo {author} {\bibfnamefont {D.~E.}\ \bibnamefont
  {Makarov}},\ }\bibfield  {title} {\enquote {\bibinfo {title} {{Tensor
  propagator for iterative quantum time evolution of reduced density matrices.
  II. Numerical methodology}},}\ }\href {https://doi.org/10.1063/1.469509}
  {\bibfield  {journal} {\bibinfo  {journal} {The Journal of Chemical Physics}\
  }\textbf {\bibinfo {volume} {102}},\ \bibinfo {pages} {4611--4618} (\bibinfo
  {year} {1995}{\natexlab{b}})}\BibitemShut {NoStop}%
\bibitem [{\citenamefont {Thorwart}, \citenamefont {Reimann},\ and\
  \citenamefont {H\"anggi}(2000)}]{Thorwart00}%
  \BibitemOpen
  \bibfield  {author} {\bibinfo {author} {\bibfnamefont {M.}~\bibnamefont
  {Thorwart}}, \bibinfo {author} {\bibfnamefont {P.}~\bibnamefont {Reimann}},\
  and\ \bibinfo {author} {\bibfnamefont {P.}~\bibnamefont {H\"anggi}},\
  }\bibfield  {title} {\enquote {\bibinfo {title} {Iterative algorithm versus
  analytic solutions of the parametrically driven dissipative quantum harmonic
  oscillator},}\ }\href {https://doi.org/10.1103/PhysRevE.62.5808} {\bibfield
  {journal} {\bibinfo  {journal} {Phys. Rev. E}\ }\textbf {\bibinfo {volume}
  {62}},\ \bibinfo {pages} {5808--5817} (\bibinfo {year} {2000})}\BibitemShut
  {NoStop}%
\bibitem [{\citenamefont {Jadhao}\ and\ \citenamefont
  {Makri}(2008)}]{JadhaoMakri08}%
  \BibitemOpen
  \bibfield  {author} {\bibinfo {author} {\bibfnamefont {V.}~\bibnamefont
  {Jadhao}}\ and\ \bibinfo {author} {\bibfnamefont {N.}~\bibnamefont {Makri}},\
  }\bibfield  {title} {\enquote {\bibinfo {title} {Iterative monte carlo for
  quantum dynamics},}\ }\href {https://doi.org/10.1063/1.3000393} {\bibfield
  {journal} {\bibinfo  {journal} {The Journal of Chemical Physics}\ }\textbf
  {\bibinfo {volume} {129}},\ \bibinfo {pages} {161102} (\bibinfo {year}
  {2008})}\BibitemShut {NoStop}%
\bibitem [{\citenamefont {Makri}(2014)}]{MakriJCP2014}%
  \BibitemOpen
  \bibfield  {author} {\bibinfo {author} {\bibfnamefont {N.}~\bibnamefont
  {Makri}},\ }\bibfield  {title} {\enquote {\bibinfo {title} {{Blip
  decomposition of the path integral: Exponential acceleration of real-time
  calculations on quantum dissipative systems}},}\ }\href
  {https://doi.org/10.1063/1.4896736} {\bibfield  {journal} {\bibinfo
  {journal} {The Journal of Chemical Physics}\ }\textbf {\bibinfo {volume}
  {141}},\ \bibinfo {pages} {134117} (\bibinfo {year} {2014})}\BibitemShut
  {NoStop}%
\bibitem [{\citenamefont {Segal}, \citenamefont {Millis},\ and\ \citenamefont
  {Reichman}(2010)}]{Reichman2010}%
  \BibitemOpen
  \bibfield  {author} {\bibinfo {author} {\bibfnamefont {D.}~\bibnamefont
  {Segal}}, \bibinfo {author} {\bibfnamefont {A.~J.}\ \bibnamefont {Millis}},\
  and\ \bibinfo {author} {\bibfnamefont {D.~R.}\ \bibnamefont {Reichman}},\
  }\bibfield  {title} {\enquote {\bibinfo {title} {Numerically exact
  path-integral simulation of nonequilibrium quantum transport and
  dissipation},}\ }\href {https://doi.org/10.1103/PhysRevB.82.205323}
  {\bibfield  {journal} {\bibinfo  {journal} {Phys. Rev. B}\ }\textbf {\bibinfo
  {volume} {82}},\ \bibinfo {pages} {205323} (\bibinfo {year}
  {2010})}\BibitemShut {NoStop}%
\bibitem [{\citenamefont {Kato}\ and\ \citenamefont
  {Tanimura}(2015)}]{KT15JCP}%
  \BibitemOpen
  \bibfield  {author} {\bibinfo {author} {\bibfnamefont {A.}~\bibnamefont
  {Kato}}\ and\ \bibinfo {author} {\bibfnamefont {Y.}~\bibnamefont
  {Tanimura}},\ }\bibfield  {title} {\enquote {\bibinfo {title} {Quantum heat
  transport of a two-qubit system: Interplay between system-bath coherence and
  qubit-qubit coherence},}\ }\href {https://doi.org/10.1063/1.4928192}
  {\bibfield  {journal} {\bibinfo  {journal} {The Journal of Chemical Physics}\
  }\textbf {\bibinfo {volume} {143}},\ \bibinfo {pages} {064107} (\bibinfo
  {year} {2015})}\BibitemShut {NoStop}%
\bibitem [{\citenamefont {Kato}\ and\ \citenamefont
  {Tanimura}(2016)}]{KT16JCP}%
  \BibitemOpen
  \bibfield  {author} {\bibinfo {author} {\bibfnamefont {A.}~\bibnamefont
  {Kato}}\ and\ \bibinfo {author} {\bibfnamefont {Y.}~\bibnamefont
  {Tanimura}},\ }\bibfield  {title} {\enquote {\bibinfo {title} {Quantum heat
  current under non-perturbative and non-\uppercase{M}arkovian conditions:
  \uppercase{A}pplications to heat machines},}\ }\href
  {https://doi.org/10.1063/1.4971370} {\bibfield  {journal} {\bibinfo
  {journal} {The Journal of Chemical Physics}\ }\textbf {\bibinfo {volume}
  {145}},\ \bibinfo {pages} {224105} (\bibinfo {year} {2016})},\ \Eprint
  {https://arxiv.org/abs/1609.08783} {arXiv:1609.08783} \BibitemShut {NoStop}%
\bibitem [{\citenamefont {Sakamoto}\ and\ \citenamefont
  {Tanimura}(2020)}]{ST20JCP}%
  \BibitemOpen
  \bibfield  {author} {\bibinfo {author} {\bibfnamefont {S.}~\bibnamefont
  {Sakamoto}}\ and\ \bibinfo {author} {\bibfnamefont {Y.}~\bibnamefont
  {Tanimura}},\ }\bibfield  {title} {\enquote {\bibinfo {title} {Numerically
  "exact" simulations of entropy production in the fully quantum regime:
  \uppercase{B}oltzmann entropy vs von \uppercase{N}eumann entropy},}\ }\href
  {https://doi.org/10.1063/5.0033664} {\bibfield  {journal} {\bibinfo
  {journal} {The Journal of Chemical Physics}\ }\textbf {\bibinfo {volume}
  {153}},\ \bibinfo {pages} {234107} (\bibinfo {year} {2020})},\ \Eprint
  {https://arxiv.org/abs/2012.09546} {arXiv:2012.09546} \BibitemShut {NoStop}%
\bibitem [{\citenamefont {Sakamoto}\ and\ \citenamefont
  {Tanimura}(2021)}]{ST21JPSJ}%
  \BibitemOpen
  \bibfield  {author} {\bibinfo {author} {\bibfnamefont {S.}~\bibnamefont
  {Sakamoto}}\ and\ \bibinfo {author} {\bibfnamefont {Y.}~\bibnamefont
  {Tanimura}},\ }\bibfield  {title} {\enquote {\bibinfo {title} {Open quantum
  dynamics theory for non-equilibrium work: \uppercase{H}ierarchical equations
  of motion approach},}\ }\href {https://doi.org/10.7566/JPSJ.90.033001}
  {\bibfield  {journal} {\bibinfo  {journal} {Journal of the Physical Society
  of Japan}\ }\textbf {\bibinfo {volume} {90}},\ \bibinfo {pages} {033001}
  (\bibinfo {year} {2021})},\ \Eprint {https://arxiv.org/abs/2101.106307}
  {arXiv:2101.106307} \BibitemShut {NoStop}%
\bibitem [{\citenamefont {Koyanagi}\ and\ \citenamefont
  {Tanimura}(2022{\natexlab{a}})}]{KT22JCP1}%
  \BibitemOpen
  \bibfield  {author} {\bibinfo {author} {\bibfnamefont {S.}~\bibnamefont
  {Koyanagi}}\ and\ \bibinfo {author} {\bibfnamefont {Y.}~\bibnamefont
  {Tanimura}},\ }\bibfield  {title} {\enquote {\bibinfo {title} {The laws of
  thermodynamics for quantum dissipative systems: \uppercase{A}
  quasi-equilibrium \uppercase{H}elmholtz energy approach},}\ }\href
  {https://doi.org/10.1063/5.0093666} {\bibfield  {journal} {\bibinfo
  {journal} {The Journal of Chemical Physics}\ }\textbf {\bibinfo {volume}
  {157}},\ \bibinfo {pages} {014104} (\bibinfo {year} {2022}{\natexlab{a}})},\
  \Eprint {https://arxiv.org/abs/2205.09487} {arXiv:2205.09487} \BibitemShut
  {NoStop}%
\bibitem [{\citenamefont {Koyanagi}\ and\ \citenamefont
  {Tanimura}(2022{\natexlab{b}})}]{KT22JCP2}%
  \BibitemOpen
  \bibfield  {author} {\bibinfo {author} {\bibfnamefont {S.}~\bibnamefont
  {Koyanagi}}\ and\ \bibinfo {author} {\bibfnamefont {Y.}~\bibnamefont
  {Tanimura}},\ }\bibfield  {title} {\enquote {\bibinfo {title} {Numerically
  “exact” simulations of a quantum carnot cycle: \uppercase{A}nalysis using
  thermodynamic work diagrams},}\ }\href {https://doi.org/10.1063/5.0107305}
  {\bibfield  {journal} {\bibinfo  {journal} {The Journal of Chemical Physics}\
  }\textbf {\bibinfo {volume} {157}},\ \bibinfo {pages} {084110} (\bibinfo
  {year} {2022}{\natexlab{b}})},\ \Eprint {https://arxiv.org/abs/2205.09487}
  {arXiv:2205.09487} \BibitemShut {NoStop}%
\bibitem [{\citenamefont {Koyanagi}\ and\ \citenamefont
  {Tanimura}(2024{\natexlab{a}})}]{KT24JCP1}%
  \BibitemOpen
  \bibfield  {author} {\bibinfo {author} {\bibfnamefont {S.}~\bibnamefont
  {Koyanagi}}\ and\ \bibinfo {author} {\bibfnamefont {Y.}~\bibnamefont
  {Tanimura}},\ }\bibfield  {title} {\enquote {\bibinfo {title} {{Classical and
  quantum thermodynamics described as a system–bath model: The dimensionless
  minimum work principle}},}\ }\href {https://doi.org/10.1063/5.0205771}
  {\bibfield  {journal} {\bibinfo  {journal} {The Journal of Chemical Physics}\
  }\textbf {\bibinfo {volume} {160}},\ \bibinfo {pages} {234112} (\bibinfo
  {year} {2024}{\natexlab{a}})},\ \Eprint {https://arxiv.org/abs/2405.16787}
  {arXiv:2405.16787} \BibitemShut {NoStop}%
\bibitem [{\citenamefont {Koyanagi}\ and\ \citenamefont
  {Tanimura}(2024{\natexlab{b}})}]{KT24JCP2}%
  \BibitemOpen
  \bibfield  {author} {\bibinfo {author} {\bibfnamefont {S.}~\bibnamefont
  {Koyanagi}}\ and\ \bibinfo {author} {\bibfnamefont {Y.}~\bibnamefont
  {Tanimura}},\ }\bibfield  {title} {\enquote {\bibinfo {title} {Classical and
  quantum thermodynamics in non-equilibrium regime: \uppercase{A}pplication to
  thermostatic \uppercase{S}tirling engine},}\ }\href
  {https://doi.org/10.1063/5.0220685} {\bibfield  {journal} {\bibinfo
  {journal} {The Journal of Chemical Physics}\ }\textbf {\bibinfo {volume}
  {161}},\ \bibinfo {pages} {114113} (\bibinfo {year} {2024}{\natexlab{b}})},\
  \Eprint {https://arxiv.org/abs/2405.17791} {arXiv:2405.17791} \BibitemShut
  {NoStop}%
\bibitem [{\citenamefont {Ullersma}(1966{\natexlab{a}})}]{Ullersma1966_1}%
  \BibitemOpen
  \bibfield  {author} {\bibinfo {author} {\bibfnamefont {P.}~\bibnamefont
  {Ullersma}},\ }\bibfield  {title} {\enquote {\bibinfo {title} {An exactly
  solvable model for \uppercase{B}rownian motion: I. derivation of the
  \uppercase{L}angevin equation},}\ }\href
  {https://doi.org/https://doi.org/10.1016/0031-8914(66)90102-9} {\bibfield
  {journal} {\bibinfo  {journal} {Physica}\ }\textbf {\bibinfo {volume} {32}},\
  \bibinfo {pages} {27--55} (\bibinfo {year} {1966}{\natexlab{a}})}\BibitemShut
  {NoStop}%
\bibitem [{\citenamefont {Ullersma}(1966{\natexlab{b}})}]{Ullersma1966_2}%
  \BibitemOpen
  \bibfield  {author} {\bibinfo {author} {\bibfnamefont {P.}~\bibnamefont
  {Ullersma}},\ }\bibfield  {title} {\enquote {\bibinfo {title} {An exactly
  solvable model for brownian motion: Ii. derivation of the
  \uppercase{F}okker-\uppercase{P}lanck equation and the master equation},}\
  }\href {https://doi.org/https://doi.org/10.1016/0031-8914(66)90103-0}
  {\bibfield  {journal} {\bibinfo  {journal} {Physica}\ }\textbf {\bibinfo
  {volume} {32}},\ \bibinfo {pages} {56--73} (\bibinfo {year}
  {1966}{\natexlab{b}})}\BibitemShut {NoStop}%
\bibitem [{\citenamefont {Caldeira}\ and\ \citenamefont
  {Leggett}(1983)}]{CALDEIRA1983587}%
  \BibitemOpen
  \bibfield  {author} {\bibinfo {author} {\bibfnamefont {A.}~\bibnamefont
  {Caldeira}}\ and\ \bibinfo {author} {\bibfnamefont {A.}~\bibnamefont
  {Leggett}},\ }\bibfield  {title} {\enquote {\bibinfo {title} {Path integral
  approach to quantum brownian motion},}\ }\href
  {https://doi.org/https://doi.org/10.1016/0378-4371(83)90013-4} {\bibfield
  {journal} {\bibinfo  {journal} {Physica A: Statistical Mechanics and its
  Applications}\ }\textbf {\bibinfo {volume} {121}},\ \bibinfo {pages}
  {587--616} (\bibinfo {year} {1983})}\BibitemShut {NoStop}%
\bibitem [{\citenamefont {Koyanagi}\ and\ \citenamefont
  {Tanimura}(2024{\natexlab{c}})}]{KT24JCP4}%
  \BibitemOpen
  \bibfield  {author} {\bibinfo {author} {\bibfnamefont {S.}~\bibnamefont
  {Koyanagi}}\ and\ \bibinfo {author} {\bibfnamefont {Y.}~\bibnamefont
  {Tanimura}},\ }\bibfield  {title} {\enquote {\bibinfo {title} {Thermodynamic
  hierarchical equations of motion and their application to \uppercase{C}arnot
  engine},}\ }\href@noop {} {\  (\bibinfo {year}
  {2024}{\natexlab{c}})}\BibitemShut {NoStop}%
\bibitem [{\citenamefont {Hu}, \citenamefont {Xu},\ and\ \citenamefont
  {Yan}(2010)}]{YanPade10A}%
  \BibitemOpen
  \bibfield  {author} {\bibinfo {author} {\bibfnamefont {J.}~\bibnamefont
  {Hu}}, \bibinfo {author} {\bibfnamefont {R.-X.}\ \bibnamefont {Xu}},\ and\
  \bibinfo {author} {\bibfnamefont {Y.}~\bibnamefont {Yan}},\ }\bibfield
  {title} {\enquote {\bibinfo {title} {{Communication: Padé spectrum
  decomposition of \uppercase{F}ermi function and \uppercase{B}ose
  function}},}\ }\href {https://doi.org/10.1063/1.3484491} {\bibfield
  {journal} {\bibinfo  {journal} {The Journal of Chemical Physics}\ }\textbf
  {\bibinfo {volume} {133}},\ \bibinfo {pages} {101106} (\bibinfo {year}
  {2010})}\BibitemShut {NoStop}%
\bibitem [{\citenamefont {Tian}\ \emph {et~al.}(2010)\citenamefont {Tian},
  \citenamefont {Ding}, \citenamefont {Xu},\ and\ \citenamefont
  {Yan}}]{YanPade10B}%
  \BibitemOpen
  \bibfield  {author} {\bibinfo {author} {\bibfnamefont {B.-L.}\ \bibnamefont
  {Tian}}, \bibinfo {author} {\bibfnamefont {J.-J.}\ \bibnamefont {Ding}},
  \bibinfo {author} {\bibfnamefont {R.-X.}\ \bibnamefont {Xu}},\ and\ \bibinfo
  {author} {\bibfnamefont {Y.}~\bibnamefont {Yan}},\ }\bibfield  {title}
  {\enquote {\bibinfo {title} {{Biexponential theory of \uppercase{D}rude
  dissipation via hierarchical quantum master equation}},}\ }\href
  {https://doi.org/10.1063/1.3491270} {\bibfield  {journal} {\bibinfo
  {journal} {The Journal of Chemical Physics}\ }\textbf {\bibinfo {volume}
  {133}},\ \bibinfo {pages} {114112} (\bibinfo {year} {2010})}\BibitemShut
  {NoStop}%
\bibitem [{\citenamefont {Hu}\ \emph {et~al.}(2011)\citenamefont {Hu},
  \citenamefont {Luo}, \citenamefont {Jiang}, \citenamefont {Xu},\ and\
  \citenamefont {Yan}}]{YanPade11}%
  \BibitemOpen
  \bibfield  {author} {\bibinfo {author} {\bibfnamefont {J.}~\bibnamefont
  {Hu}}, \bibinfo {author} {\bibfnamefont {M.}~\bibnamefont {Luo}}, \bibinfo
  {author} {\bibfnamefont {F.}~\bibnamefont {Jiang}}, \bibinfo {author}
  {\bibfnamefont {R.-X.}\ \bibnamefont {Xu}},\ and\ \bibinfo {author}
  {\bibfnamefont {Y.}~\bibnamefont {Yan}},\ }\bibfield  {title} {\enquote
  {\bibinfo {title} {{\uppercase{P}adé spectrum decompositions of quantum
  distribution functions and optimal hierarchical equations of motion
  construction for quantum open systems}},}\ }\href
  {https://doi.org/10.1063/1.3602466} {\bibfield  {journal} {\bibinfo
  {journal} {The Journal of Chemical Physics}\ }\textbf {\bibinfo {volume}
  {134}},\ \bibinfo {pages} {244106} (\bibinfo {year} {2011})}\BibitemShut
  {NoStop}%
\bibitem [{\citenamefont {Sakurai}\ and\ \citenamefont
  {Tanimura}(2013)}]{ST13JPSJ}%
  \BibitemOpen
  \bibfield  {author} {\bibinfo {author} {\bibfnamefont {A.}~\bibnamefont
  {Sakurai}}\ and\ \bibinfo {author} {\bibfnamefont {Y.}~\bibnamefont
  {Tanimura}},\ }\bibfield  {title} {\enquote {\bibinfo {title} {An approach to
  quantum transport based on reduced hierarchy equations of motion: Application
  to a resonant tunneling diode},}\ }\href
  {https://doi.org/10.7566/JPSJ.82.033707} {\bibfield  {journal} {\bibinfo
  {journal} {Journal of the Physical Society of Japan}\ }\textbf {\bibinfo
  {volume} {82}},\ \bibinfo {pages} {033707} (\bibinfo {year}
  {2013})}\BibitemShut {NoStop}%
\bibitem [{\citenamefont {Sakurai}\ and\ \citenamefont
  {Tanimura}(2014)}]{ST14NJP}%
  \BibitemOpen
  \bibfield  {author} {\bibinfo {author} {\bibfnamefont {A.}~\bibnamefont
  {Sakurai}}\ and\ \bibinfo {author} {\bibfnamefont {Y.}~\bibnamefont
  {Tanimura}},\ }\bibfield  {title} {\enquote {\bibinfo {title} {Self-excited
  current oscillations in a resonant tunneling diode described by a model based
  on the \uppercase{C}aldeira{\textendash}\uppercase{L}eggett
  \uppercase{H}amiltonian},}\ }\href
  {https://doi.org/10.1088/1367-2630/16/1/015002} {\bibfield  {journal}
  {\bibinfo  {journal} {New Journal of Physics}\ }\textbf {\bibinfo {volume}
  {16}},\ \bibinfo {pages} {015002} (\bibinfo {year} {2014})}\BibitemShut
  {NoStop}%
\bibitem [{\citenamefont {Grossmann}, \citenamefont {Sakurai},\ and\
  \citenamefont {Tanimura}(2016)}]{GST16JPSJ}%
  \BibitemOpen
  \bibfield  {author} {\bibinfo {author} {\bibfnamefont {F.}~\bibnamefont
  {Grossmann}}, \bibinfo {author} {\bibfnamefont {A.}~\bibnamefont {Sakurai}},\
  and\ \bibinfo {author} {\bibfnamefont {Y.}~\bibnamefont {Tanimura}},\
  }\bibfield  {title} {\enquote {\bibinfo {title} {Electron pumping under
  non-\uppercase{M}arkovian dissipation: The role of the self-consistent
  field},}\ }\href {https://doi.org/10.7566/JPSJ.85.034803} {\bibfield
  {journal} {\bibinfo  {journal} {Journal of the Physical Society of Japan}\
  }\textbf {\bibinfo {volume} {85}},\ \bibinfo {pages} {034803} (\bibinfo
  {year} {2016})}\BibitemShut {NoStop}%
\bibitem [{\citenamefont {Ikeda}, \citenamefont {Dijkstra},\ and\ \citenamefont
  {Tanimura}(2019)}]{IDT19JCP}%
  \BibitemOpen
  \bibfield  {author} {\bibinfo {author} {\bibfnamefont {T.}~\bibnamefont
  {Ikeda}}, \bibinfo {author} {\bibfnamefont {A.~G.}\ \bibnamefont
  {Dijkstra}},\ and\ \bibinfo {author} {\bibfnamefont {Y.}~\bibnamefont
  {Tanimura}},\ }\bibfield  {title} {\enquote {\bibinfo {title} {Modeling and
  analyzing a photo-driven molecular motor system: \uppercase{R}atchet dynamics
  and non-linear optical spectra},}\ }\href {https://doi.org/10.1063/1.5086948}
  {\bibfield  {journal} {\bibinfo  {journal} {The Journal of Chemical Physics}\
  }\textbf {\bibinfo {volume} {150}},\ \bibinfo {pages} {114103} (\bibinfo
  {year} {2019})}\BibitemShut {NoStop}%
\bibitem [{\citenamefont {Frensley}(1990)}]{Frensley1990}%
  \BibitemOpen
  \bibfield  {author} {\bibinfo {author} {\bibfnamefont {W.~R.}\ \bibnamefont
  {Frensley}},\ }\bibfield  {title} {\enquote {\bibinfo {title} {Boundary
  conditions for open quantum systems driven far from equilibrium},}\ }\href
  {https://doi.org/10.1103/RevModPhys.62.745} {\bibfield  {journal} {\bibinfo
  {journal} {Rev. Mod. Phys.}\ }\textbf {\bibinfo {volume} {62}},\ \bibinfo
  {pages} {745--791} (\bibinfo {year} {1990})}\BibitemShut {NoStop}%
\bibitem [{\citenamefont {Risken}\ and\ \citenamefont
  {Risken}(1996)}]{risken1996fokker}%
  \BibitemOpen
  \bibfield  {author} {\bibinfo {author} {\bibfnamefont {H.}~\bibnamefont
  {Risken}}\ and\ \bibinfo {author} {\bibfnamefont {H.}~\bibnamefont
  {Risken}},\ }\href@noop {} {\emph {\bibinfo {title}
  {Fokker-\uppercase{P}lanck equation}}}\ (\bibinfo  {publisher} {Springer},\
  \bibinfo {year} {1996})\BibitemShut {NoStop}%
\bibitem [{\citenamefont {Chen}\ \emph {et~al.}(2022)\citenamefont {Chen},
  \citenamefont {Wang}, \citenamefont {Zheng}, \citenamefont {Xu},\ and\
  \citenamefont {Yan}}]{10.1063/5.0095961}%
  \BibitemOpen
  \bibfield  {author} {\bibinfo {author} {\bibfnamefont {Z.-H.}\ \bibnamefont
  {Chen}}, \bibinfo {author} {\bibfnamefont {Y.}~\bibnamefont {Wang}}, \bibinfo
  {author} {\bibfnamefont {X.}~\bibnamefont {Zheng}}, \bibinfo {author}
  {\bibfnamefont {R.-X.}\ \bibnamefont {Xu}},\ and\ \bibinfo {author}
  {\bibfnamefont {Y.}~\bibnamefont {Yan}},\ }\bibfield  {title} {\enquote
  {\bibinfo {title} {{Universal time-domain \uppercase{P}rony fitting
  decomposition for optimized hierarchical quantum master equations}},}\ }\href
  {https://doi.org/10.1063/5.0095961} {\bibfield  {journal} {\bibinfo
  {journal} {The Journal of Chemical Physics}\ }\textbf {\bibinfo {volume}
  {156}},\ \bibinfo {pages} {221102} (\bibinfo {year} {2022})},\ \Eprint
  {https://arxiv.org/abs/https://pubs.aip.org/aip/jcp/article-pdf/doi/10.1063/5.0095961/16543316/221102\_1\_online.pdf}
  {https://pubs.aip.org/aip/jcp/article-pdf/doi/10.1063/5.0095961/16543316/221102\_1\_online.pdf}
  \BibitemShut {NoStop}%
\bibitem [{\citenamefont {Xu}\ \emph {et~al.}(2022)\citenamefont {Xu},
  \citenamefont {Yan}, \citenamefont {Shi}, \citenamefont {Ankerhold},\ and\
  \citenamefont {Stockburger}}]{PhysRevLett.129.230601}%
  \BibitemOpen
  \bibfield  {author} {\bibinfo {author} {\bibfnamefont {M.}~\bibnamefont
  {Xu}}, \bibinfo {author} {\bibfnamefont {Y.}~\bibnamefont {Yan}}, \bibinfo
  {author} {\bibfnamefont {Q.}~\bibnamefont {Shi}}, \bibinfo {author}
  {\bibfnamefont {J.}~\bibnamefont {Ankerhold}},\ and\ \bibinfo {author}
  {\bibfnamefont {J.~T.}\ \bibnamefont {Stockburger}},\ }\bibfield  {title}
  {\enquote {\bibinfo {title} {Taming quantum noise for efficient low
  temperature simulations of open quantum systems},}\ }\href
  {https://doi.org/10.1103/PhysRevLett.129.230601} {\bibfield  {journal}
  {\bibinfo  {journal} {Phys. Rev. Lett.}\ }\textbf {\bibinfo {volume} {129}},\
  \bibinfo {pages} {230601} (\bibinfo {year} {2022})}\BibitemShut {NoStop}%
\bibitem [{\citenamefont {Kramers}(1940)}]{KRAMERS1940284}%
  \BibitemOpen
  \bibfield  {author} {\bibinfo {author} {\bibfnamefont {H.}~\bibnamefont
  {Kramers}},\ }\bibfield  {title} {\enquote {\bibinfo {title} {Brownian motion
  in a field of force and the diffusion model of chemical reactions},}\ }\href
  {https://doi.org/https://doi.org/10.1016/S0031-8914(40)90098-2} {\bibfield
  {journal} {\bibinfo  {journal} {Physica}\ }\textbf {\bibinfo {volume} {7}},\
  \bibinfo {pages} {284--304} (\bibinfo {year} {1940})}\BibitemShut {NoStop}%
\bibitem [{\citenamefont {Tanimura}\ and\ \citenamefont
  {Maruyama}(1997)}]{TM97JCP}%
  \BibitemOpen
  \bibfield  {author} {\bibinfo {author} {\bibfnamefont {Y.}~\bibnamefont
  {Tanimura}}\ and\ \bibinfo {author} {\bibfnamefont {Y.}~\bibnamefont
  {Maruyama}},\ }\bibfield  {title} {\enquote {\bibinfo {title}
  {Gaussian-\uppercase{M}arkovian quantum \uppercase{F}okker-\uppercase{P}lanck
  approach to nonlinear spectroscopy of a displaced \uppercase{M}orse
  potentials system: Dissociation, predissociation, and optical
  \uppercase{S}tark effects},}\ }\href {https://doi.org/10.1063/1.474531}
  {\bibfield  {journal} {\bibinfo  {journal} {The Journal of Chemical Physics}\
  }\textbf {\bibinfo {volume} {107}},\ \bibinfo {pages} {1779--1793} (\bibinfo
  {year} {1997})}\BibitemShut {NoStop}%
\bibitem [{\citenamefont {Lindblad}(1976)}]{Lindblad1976}%
  \BibitemOpen
  \bibfield  {author} {\bibinfo {author} {\bibfnamefont {G.}~\bibnamefont
  {Lindblad}},\ }\bibfield  {title} {\enquote {\bibinfo {title} {On the
  generators of quantum dynamical semigroups},}\ }\href
  {https://doi.org/10.1007/BF01608499} {\bibfield  {journal} {\bibinfo
  {journal} {Communications in Mathematical Physics}\ }\textbf {\bibinfo
  {volume} {48}},\ \bibinfo {pages} {119--130} (\bibinfo {year}
  {1976})}\BibitemShut {NoStop}%
\bibitem [{\citenamefont {Esposito}, \citenamefont {Harbola},\ and\
  \citenamefont {Mukamel}(2009)}]{RevModPhys.81.1665}%
  \BibitemOpen
  \bibfield  {author} {\bibinfo {author} {\bibfnamefont {M.}~\bibnamefont
  {Esposito}}, \bibinfo {author} {\bibfnamefont {U.}~\bibnamefont {Harbola}},\
  and\ \bibinfo {author} {\bibfnamefont {S.}~\bibnamefont {Mukamel}},\
  }\bibfield  {title} {\enquote {\bibinfo {title} {Nonequilibrium fluctuations,
  fluctuation theorems, and counting statistics in quantum systems},}\ }\href
  {https://doi.org/10.1103/RevModPhys.81.1665} {\bibfield  {journal} {\bibinfo
  {journal} {Rev. Mod. Phys.}\ }\textbf {\bibinfo {volume} {81}},\ \bibinfo
  {pages} {1665--1702} (\bibinfo {year} {2009})}\BibitemShut {NoStop}%
\bibitem [{\citenamefont {Campisi}, \citenamefont {H\"anggi},\ and\
  \citenamefont {Talkner}(2011)}]{RevModPhys.83.771}%
  \BibitemOpen
  \bibfield  {author} {\bibinfo {author} {\bibfnamefont {M.}~\bibnamefont
  {Campisi}}, \bibinfo {author} {\bibfnamefont {P.}~\bibnamefont {H\"anggi}},\
  and\ \bibinfo {author} {\bibfnamefont {P.}~\bibnamefont {Talkner}},\
  }\bibfield  {title} {\enquote {\bibinfo {title} {Colloquium: Quantum
  fluctuation relations: Foundations and applications},}\ }\href
  {https://doi.org/10.1103/RevModPhys.83.771} {\bibfield  {journal} {\bibinfo
  {journal} {Rev. Mod. Phys.}\ }\textbf {\bibinfo {volume} {83}},\ \bibinfo
  {pages} {771--791} (\bibinfo {year} {2011})}\BibitemShut {NoStop}%
\bibitem [{\citenamefont {Sagawa}(2023)}]{sagawa20232law}%
  \BibitemOpen
  \bibfield  {author} {\bibinfo {author} {\bibfnamefont {T.}~\bibnamefont
  {Sagawa}},\ }\bibfield  {title} {\enquote {\bibinfo {title} {Second law-like
  inequalities with quantum relative entropy: An introduction},}\ }\href
  {https://arxiv.org/abs/1202.0983} {\  (\bibinfo {year} {2023})},\ \Eprint
  {https://arxiv.org/abs/1202.0983} {arXiv:1202.0983 [cond-mat.stat-mech]}
  \BibitemShut {NoStop}%
\bibitem [{\citenamefont {Schmiedl}\ and\ \citenamefont
  {Seifert}(2007)}]{Schmiedl_2008}%
  \BibitemOpen
  \bibfield  {author} {\bibinfo {author} {\bibfnamefont {T.}~\bibnamefont
  {Schmiedl}}\ and\ \bibinfo {author} {\bibfnamefont {U.}~\bibnamefont
  {Seifert}},\ }\bibfield  {title} {\enquote {\bibinfo {title} {Efficiency at
  maximum power: An analytically solvable model for stochastic heat engines},}\
  }\href {https://doi.org/10.1209/0295-5075/81/20003} {\bibfield  {journal}
  {\bibinfo  {journal} {Europhysics Letters}\ }\textbf {\bibinfo {volume}
  {81}},\ \bibinfo {pages} {20003} (\bibinfo {year} {2007})}\BibitemShut
  {NoStop}%
\bibitem [{\citenamefont {Seifert}(2012)}]{Stochastic_Seifert_2012}%
  \BibitemOpen
  \bibfield  {author} {\bibinfo {author} {\bibfnamefont {U.}~\bibnamefont
  {Seifert}},\ }\bibfield  {title} {\enquote {\bibinfo {title} {Stochastic
  thermodynamics, fluctuation theorems and molecular machines},}\ }\href
  {https://doi.org/10.1088/0034-4885/75/12/126001} {\bibfield  {journal}
  {\bibinfo  {journal} {Reports on Progress in Physics}\ }\textbf {\bibinfo
  {volume} {75}},\ \bibinfo {pages} {126001} (\bibinfo {year}
  {2012})}\BibitemShut {NoStop}%
\bibitem [{\citenamefont {Esposito}\ and\ \citenamefont {den
  Broeck}(2011)}]{Esposito2011}%
  \BibitemOpen
  \bibfield  {author} {\bibinfo {author} {\bibfnamefont {M.}~\bibnamefont
  {Esposito}}\ and\ \bibinfo {author} {\bibfnamefont {C.~V.}\ \bibnamefont {den
  Broeck}},\ }\bibfield  {title} {\enquote {\bibinfo {title} {Second law and
  landauer principle far from equilibrium},}\ }\href
  {https://doi.org/10.1209/0295-5075/95/40004} {\bibfield  {journal} {\bibinfo
  {journal} {Europhysics Letters}\ }\textbf {\bibinfo {volume} {95}},\ \bibinfo
  {pages} {40004} (\bibinfo {year} {2011})}\BibitemShut {NoStop}%
\bibitem [{\citenamefont {Strasberg}(2022)}]{strasberg2022quantum}%
  \BibitemOpen
  \bibfield  {author} {\bibinfo {author} {\bibfnamefont {P.}~\bibnamefont
  {Strasberg}},\ }\href@noop {} {\emph {\bibinfo {title} {Quantum Stochastic
  Thermodynamics: Foundations and Selected Applications}}}\ (\bibinfo
  {publisher} {Oxford University Press},\ \bibinfo {year} {2022})\BibitemShut
  {NoStop}%
\bibitem [{\citenamefont {Rao}\ and\ \citenamefont
  {Esposito}(2018)}]{RaoEsposit2018}%
  \BibitemOpen
  \bibfield  {author} {\bibinfo {author} {\bibfnamefont {R.}~\bibnamefont
  {Rao}}\ and\ \bibinfo {author} {\bibfnamefont {M.}~\bibnamefont {Esposito}},\
  }\bibfield  {title} {\enquote {\bibinfo {title} {Conservation laws shape
  dissipation},}\ }\href {https://doi.org/10.1088/1367-2630/aaa15f} {\bibfield
  {journal} {\bibinfo  {journal} {New Journal of Physics}\ }\textbf {\bibinfo
  {volume} {20}},\ \bibinfo {pages} {023007} (\bibinfo {year}
  {2018})}\BibitemShut {NoStop}%
\bibitem [{\citenamefont {Cockrell}\ and\ \citenamefont
  {Ford}(2022)}]{PhysRevE.105.064124}%
  \BibitemOpen
  \bibfield  {author} {\bibinfo {author} {\bibfnamefont {C.}~\bibnamefont
  {Cockrell}}\ and\ \bibinfo {author} {\bibfnamefont {I.~J.}\ \bibnamefont
  {Ford}},\ }\bibfield  {title} {\enquote {\bibinfo {title} {Stochastic
  thermodynamics in a non-\uppercase{M}arkovian dynamical system},}\ }\href
  {https://doi.org/10.1103/PhysRevE.105.064124} {\bibfield  {journal} {\bibinfo
   {journal} {Phys. Rev. E}\ }\textbf {\bibinfo {volume} {105}},\ \bibinfo
  {pages} {064124} (\bibinfo {year} {2022})}\BibitemShut {NoStop}%
\bibitem [{\citenamefont {Wu}\ \emph {et~al.}(1998)\citenamefont {Wu},
  \citenamefont {Chen}, \citenamefont {Sun}, \citenamefont {Wu},\ and\
  \citenamefont {Zhu}}]{wu1998performance}%
  \BibitemOpen
  \bibfield  {author} {\bibinfo {author} {\bibfnamefont {F.}~\bibnamefont
  {Wu}}, \bibinfo {author} {\bibfnamefont {L.}~\bibnamefont {Chen}}, \bibinfo
  {author} {\bibfnamefont {F.}~\bibnamefont {Sun}}, \bibinfo {author}
  {\bibfnamefont {C.}~\bibnamefont {Wu}},\ and\ \bibinfo {author}
  {\bibfnamefont {Y.}~\bibnamefont {Zhu}},\ }\bibfield  {title} {\enquote
  {\bibinfo {title} {Performance and optimization criteria for forward and
  reverse quantum \uppercase{S}tirling cycles},}\ }\href
  {https://doi.org/10.1016/S0196-8904(97)10037-1} {\bibfield  {journal}
  {\bibinfo  {journal} {Energy conversion and management}\ }\textbf {\bibinfo
  {volume} {39}},\ \bibinfo {pages} {733--739} (\bibinfo {year}
  {1998})}\BibitemShut {NoStop}%
\bibitem [{\citenamefont {Raja}\ \emph {et~al.}(2021)\citenamefont {Raja},
  \citenamefont {Maniscalco}, \citenamefont {Paraoanu}, \citenamefont
  {Pekola},\ and\ \citenamefont {Gullo}}]{Stirling2021}%
  \BibitemOpen
  \bibfield  {author} {\bibinfo {author} {\bibfnamefont {S.~H.}\ \bibnamefont
  {Raja}}, \bibinfo {author} {\bibfnamefont {S.}~\bibnamefont {Maniscalco}},
  \bibinfo {author} {\bibfnamefont {G.~S.}\ \bibnamefont {Paraoanu}}, \bibinfo
  {author} {\bibfnamefont {J.~P.}\ \bibnamefont {Pekola}},\ and\ \bibinfo
  {author} {\bibfnamefont {N.~L.}\ \bibnamefont {Gullo}},\ }\bibfield  {title}
  {\enquote {\bibinfo {title} {Finite-time quantum \uppercase{S}tirling heat
  engine},}\ }\href {https://doi.org/10.1088/1367-2630/abe9d7} {\bibfield
  {journal} {\bibinfo  {journal} {New Journal of Physics}\ }\textbf {\bibinfo
  {volume} {23}},\ \bibinfo {pages} {033034} (\bibinfo {year}
  {2021})}\BibitemShut {NoStop}%
\bibitem [{\citenamefont {Huang}\ \emph {et~al.}(2014)\citenamefont {Huang},
  \citenamefont {Niu}, \citenamefont {Xiu},\ and\ \citenamefont
  {Yi}}]{huang2014quantum}%
  \BibitemOpen
  \bibfield  {author} {\bibinfo {author} {\bibfnamefont {X.-L.}\ \bibnamefont
  {Huang}}, \bibinfo {author} {\bibfnamefont {X.-Y.}\ \bibnamefont {Niu}},
  \bibinfo {author} {\bibfnamefont {X.-M.}\ \bibnamefont {Xiu}},\ and\ \bibinfo
  {author} {\bibfnamefont {X.-X.}\ \bibnamefont {Yi}},\ }\bibfield  {title}
  {\enquote {\bibinfo {title} {Quantum stirling heat engine and refrigerator
  with single and coupled spin systems},}\ }\href
  {https://doi.org/10.1140/epjd/e2013-40536-0} {\bibfield  {journal} {\bibinfo
  {journal} {The European Physical Journal D}\ }\textbf {\bibinfo {volume}
  {68}},\ \bibinfo {pages} {1--8} (\bibinfo {year} {2014})}\BibitemShut
  {NoStop}%
\bibitem [{\citenamefont {Xia}\ \emph {et~al.}(2024)\citenamefont {Xia},
  \citenamefont {Lv}, \citenamefont {Pan}, \citenamefont {Chen},\ and\
  \citenamefont {Su}}]{xia2024performance}%
  \BibitemOpen
  \bibfield  {author} {\bibinfo {author} {\bibfnamefont {S.}~\bibnamefont
  {Xia}}, \bibinfo {author} {\bibfnamefont {M.}~\bibnamefont {Lv}}, \bibinfo
  {author} {\bibfnamefont {Y.}~\bibnamefont {Pan}}, \bibinfo {author}
  {\bibfnamefont {J.}~\bibnamefont {Chen}},\ and\ \bibinfo {author}
  {\bibfnamefont {S.}~\bibnamefont {Su}},\ }\bibfield  {title} {\enquote
  {\bibinfo {title} {Performance improvement of a fractional quantum stirling
  heat engine},}\ }\href {https://doi.org/10.1063/5.0187666} {\bibfield
  {journal} {\bibinfo  {journal} {Journal of Applied Physics}\ }\textbf
  {\bibinfo {volume} {135}} (\bibinfo {year} {2024})}\BibitemShut {NoStop}%
\bibitem [{\citenamefont {Shi}, \citenamefont {Zhu},\ and\ \citenamefont
  {Chen}(2011)}]{Shi2011PT}%
  \BibitemOpen
  \bibfield  {author} {\bibinfo {author} {\bibfnamefont {Q.}~\bibnamefont
  {Shi}}, \bibinfo {author} {\bibfnamefont {L.}~\bibnamefont {Zhu}},\ and\
  \bibinfo {author} {\bibfnamefont {L.}~\bibnamefont {Chen}},\ }\bibfield
  {title} {\enquote {\bibinfo {title} {{Quantum rate dynamics for proton
  transfer reaction in a model system: Effect of the rate promoting vibrational
  mode}},}\ }\href {https://doi.org/10.1063/1.3611050} {\bibfield  {journal}
  {\bibinfo  {journal} {The Journal of Chemical Physics}\ }\textbf {\bibinfo
  {volume} {135}},\ \bibinfo {pages} {044505} (\bibinfo {year} {2011})},\
  \Eprint
  {https://arxiv.org/abs/https://pubs.aip.org/aip/jcp/article-pdf/doi/10.1063/1.3611050/14817375/044505\_1\_online.pdf}
  {https://pubs.aip.org/aip/jcp/article-pdf/doi/10.1063/1.3611050/14817375/044505\_1\_online.pdf}
  \BibitemShut {NoStop}%
\bibitem [{\citenamefont {Zhang}, \citenamefont {Borrelli},\ and\ \citenamefont
  {Tanimura}(2020)}]{ZBT20JCP}%
  \BibitemOpen
  \bibfield  {author} {\bibinfo {author} {\bibfnamefont {J.}~\bibnamefont
  {Zhang}}, \bibinfo {author} {\bibfnamefont {R.}~\bibnamefont {Borrelli}},\
  and\ \bibinfo {author} {\bibfnamefont {Y.}~\bibnamefont {Tanimura}},\
  }\bibfield  {title} {\enquote {\bibinfo {title} {Proton tunneling in a
  two-dimensional potential energy surface with a non-linear system-bath
  interaction: Thermal suppression of reaction rate},}\ }\href
  {https://doi.org/10.1063/5.0010580} {\bibfield  {journal} {\bibinfo
  {journal} {The Journal of Chemical Physics}\ }\textbf {\bibinfo {volume}
  {152}},\ \bibinfo {pages} {214114} (\bibinfo {year} {2020})}\BibitemShut
  {NoStop}%
\bibitem [{\citenamefont {Ikeda}\ and\ \citenamefont
  {Tanimura}(2017)}]{IT17JCP}%
  \BibitemOpen
  \bibfield  {author} {\bibinfo {author} {\bibfnamefont {T.}~\bibnamefont
  {Ikeda}}\ and\ \bibinfo {author} {\bibfnamefont {Y.}~\bibnamefont
  {Tanimura}},\ }\bibfield  {title} {\enquote {\bibinfo {title} {Probing
  photoisomerization processes by means of multi-dimensional electronic
  spectroscopy: The multi-state quantum hierarchical
  \uppercase{F}okker-\uppercase{P}lanck equation approach},}\ }\href
  {https://doi.org/10.1063/1.4989537} {\bibfield  {journal} {\bibinfo
  {journal} {The Journal of Chemical Physics}\ }\textbf {\bibinfo {volume}
  {147}},\ \bibinfo {pages} {014102} (\bibinfo {year} {2017})}\BibitemShut
  {NoStop}%
\bibitem [{\citenamefont {Ikeda}, \citenamefont {Ito},\ and\ \citenamefont
  {Tanimura}(2015)}]{IIT15JCP}%
  \BibitemOpen
  \bibfield  {author} {\bibinfo {author} {\bibfnamefont {T.}~\bibnamefont
  {Ikeda}}, \bibinfo {author} {\bibfnamefont {H.}~\bibnamefont {Ito}},\ and\
  \bibinfo {author} {\bibfnamefont {Y.}~\bibnamefont {Tanimura}},\ }\bibfield
  {title} {\enquote {\bibinfo {title} {Analysis of 2\uppercase{D}
  \uppercase{TH}z-\uppercase{R}aman spectroscopy using a
  non-\uppercase{M}arkovian \uppercase{B}rownian oscillator model with
  nonlinear system-bath interactions},}\ }\href
  {https://doi.org/10.1063/1.4917033} {\bibfield  {journal} {\bibinfo
  {journal} {The Journal of Chemical Physics}\ }\textbf {\bibinfo {volume}
  {142}},\ \bibinfo {pages} {212421} (\bibinfo {year} {2015})}\BibitemShut
  {NoStop}%
\bibitem [{\citenamefont {Ito}\ and\ \citenamefont {Tanimura}(2016)}]{IT16JCP}%
  \BibitemOpen
  \bibfield  {author} {\bibinfo {author} {\bibfnamefont {H.}~\bibnamefont
  {Ito}}\ and\ \bibinfo {author} {\bibfnamefont {Y.}~\bibnamefont {Tanimura}},\
  }\bibfield  {title} {\enquote {\bibinfo {title} {Simulating two-dimensional
  infrared-\uppercase{R}aman and \uppercase{R}aman spectroscopies for
  intermolecular and intramolecular modes of liquid water},}\ }\href
  {https://doi.org/10.1063/1.4941842} {\bibfield  {journal} {\bibinfo
  {journal} {The Journal of Chemical Physics}\ }\textbf {\bibinfo {volume}
  {144}},\ \bibinfo {pages} {074201} (\bibinfo {year} {2016})}\BibitemShut
  {NoStop}%
\bibitem [{\citenamefont {Takahashi}\ and\ \citenamefont
  {Tanimura}(2023{\natexlab{a}})}]{TT23JCP1}%
  \BibitemOpen
  \bibfield  {author} {\bibinfo {author} {\bibfnamefont {H.}~\bibnamefont
  {Takahashi}}\ and\ \bibinfo {author} {\bibfnamefont {Y.}~\bibnamefont
  {Tanimura}},\ }\bibfield  {title} {\enquote {\bibinfo {title} {Discretized
  hierarchal equations of motion in mixed
  \uppercase{L}iouville--\uppercase{W}igner space for two-dimensional
  vibrational spectroscopies of water},}\ }\href
  {https://doi.org/10.1063/5.0135725} {\bibfield  {journal} {\bibinfo
  {journal} {The Journal of Chemical Physics}\ }\textbf {\bibinfo {volume}
  {158}},\ \bibinfo {pages} {044115} (\bibinfo {year} {2023}{\natexlab{a}})},\
  \Eprint {https://arxiv.org/abs/2302.09799} {arXiv:2302.09799} \BibitemShut
  {NoStop}%
\bibitem [{\citenamefont {Takahashi}\ and\ \citenamefont
  {Tanimura}(2023{\natexlab{b}})}]{TT23JCP2}%
  \BibitemOpen
  \bibfield  {author} {\bibinfo {author} {\bibfnamefont {H.}~\bibnamefont
  {Takahashi}}\ and\ \bibinfo {author} {\bibfnamefont {Y.}~\bibnamefont
  {Tanimura}},\ }\bibfield  {title} {\enquote {\bibinfo {title} {{Simulating
  two-dimensional correlation spectroscopies with third-order infrared and
  fifth-order infrared–\uppercase{R}aman processes of liquid water}},}\
  }\href {https://doi.org/10.1063/5.0141181} {\bibfield  {journal} {\bibinfo
  {journal} {The Journal of Chemical Physics}\ }\textbf {\bibinfo {volume}
  {158}},\ \bibinfo {pages} {124108} (\bibinfo {year} {2023}{\natexlab{b}})},\
  \Eprint {https://arxiv.org/abs/2302.09760} {arXiv:2302.09760} \BibitemShut
  {NoStop}%
\bibitem [{\citenamefont {Ikeda}\ and\ \citenamefont
  {Scholes}(2020)}]{10.1063/5.0007327}%
  \BibitemOpen
  \bibfield  {author} {\bibinfo {author} {\bibfnamefont {T.}~\bibnamefont
  {Ikeda}}\ and\ \bibinfo {author} {\bibfnamefont {G.~D.}\ \bibnamefont
  {Scholes}},\ }\bibfield  {title} {\enquote {\bibinfo {title} {{Generalization
  of the hierarchical equations of motion theory for efficient calculations
  with arbitrary correlation functions}},}\ }\href
  {https://doi.org/10.1063/5.0007327} {\bibfield  {journal} {\bibinfo
  {journal} {The Journal of Chemical Physics}\ }\textbf {\bibinfo {volume}
  {152}},\ \bibinfo {pages} {204101} (\bibinfo {year} {2020})}\BibitemShut
  {NoStop}%
\bibitem [{\citenamefont {Takahashi}\ \emph {et~al.}(2024)\citenamefont
  {Takahashi}, \citenamefont {Rudge}, \citenamefont {Kaspar}, \citenamefont
  {Thoss},\ and\ \citenamefont {Borrelli}}]{10.1063/5.0209348}%
  \BibitemOpen
  \bibfield  {author} {\bibinfo {author} {\bibfnamefont {H.}~\bibnamefont
  {Takahashi}}, \bibinfo {author} {\bibfnamefont {S.}~\bibnamefont {Rudge}},
  \bibinfo {author} {\bibfnamefont {C.}~\bibnamefont {Kaspar}}, \bibinfo
  {author} {\bibfnamefont {M.}~\bibnamefont {Thoss}},\ and\ \bibinfo {author}
  {\bibfnamefont {R.}~\bibnamefont {Borrelli}},\ }\bibfield  {title} {\enquote
  {\bibinfo {title} {{High accuracy exponential decomposition of bath
  correlation functions for arbitrary and structured spectral densities:
  Emerging methodologies and new approaches}},}\ }\href
  {https://doi.org/10.1063/5.0209348} {\bibfield  {journal} {\bibinfo
  {journal} {The Journal of Chemical Physics}\ }\textbf {\bibinfo {volume}
  {160}},\ \bibinfo {pages} {204105} (\bibinfo {year} {2024})}\BibitemShut
  {NoStop}%
\bibitem [{\citenamefont {Le~Dé}\ \emph {et~al.}(2024)\citenamefont {Le~Dé},
  \citenamefont {Jaouadi}, \citenamefont {Mangaud}, \citenamefont {Chin},\ and\
  \citenamefont {Desouter-Lecomte}}]{10.1063/5.0214051}%
  \BibitemOpen
  \bibfield  {author} {\bibinfo {author} {\bibfnamefont {B.}~\bibnamefont
  {Le~Dé}}, \bibinfo {author} {\bibfnamefont {A.}~\bibnamefont {Jaouadi}},
  \bibinfo {author} {\bibfnamefont {E.}~\bibnamefont {Mangaud}}, \bibinfo
  {author} {\bibfnamefont {A.~W.}\ \bibnamefont {Chin}},\ and\ \bibinfo
  {author} {\bibfnamefont {M.}~\bibnamefont {Desouter-Lecomte}},\ }\bibfield
  {title} {\enquote {\bibinfo {title} {{Managing temperature in open quantum
  systems strongly coupled with structured environments}},}\ }\href
  {https://doi.org/10.1063/5.0214051} {\bibfield  {journal} {\bibinfo
  {journal} {The Journal of Chemical Physics}\ }\textbf {\bibinfo {volume}
  {160}},\ \bibinfo {pages} {244102} (\bibinfo {year} {2024})}\BibitemShut
  {NoStop}%
\bibitem [{\citenamefont {Ikeda}\ and\ \citenamefont
  {Tanimura}(2018)}]{IT18CP}%
  \BibitemOpen
  \bibfield  {author} {\bibinfo {author} {\bibfnamefont {T.}~\bibnamefont
  {Ikeda}}\ and\ \bibinfo {author} {\bibfnamefont {Y.}~\bibnamefont
  {Tanimura}},\ }\bibfield  {title} {\enquote {\bibinfo {title} {Phase-space
  wavepacket dynamics of internal conversion via conical intersection:
  Multi-state quantum \uppercase{F}okker-\uppercase{P}lanck equation
  approach},}\ }\href
  {https://doi.org/https://doi.org/10.1016/j.chemphys.2018.07.013} {\bibfield
  {journal} {\bibinfo  {journal} {Chemical Physics}\ }\textbf {\bibinfo
  {volume} {515}},\ \bibinfo {pages} {203--213} (\bibinfo {year}
  {2018})}\BibitemShut {NoStop}%
\bibitem [{\citenamefont {Hou}\ \emph {et~al.}(2015)\citenamefont {Hou},
  \citenamefont {Wang}, \citenamefont {Wang}, \citenamefont {Ye}, \citenamefont
  {Xu}, \citenamefont {Zheng},\ and\ \citenamefont
  {Yan}}]{YanOptimalBasis2015}%
  \BibitemOpen
  \bibfield  {author} {\bibinfo {author} {\bibfnamefont {D.}~\bibnamefont
  {Hou}}, \bibinfo {author} {\bibfnamefont {S.}~\bibnamefont {Wang}}, \bibinfo
  {author} {\bibfnamefont {R.}~\bibnamefont {Wang}}, \bibinfo {author}
  {\bibfnamefont {L.}~\bibnamefont {Ye}}, \bibinfo {author} {\bibfnamefont
  {R.}~\bibnamefont {Xu}}, \bibinfo {author} {\bibfnamefont {X.}~\bibnamefont
  {Zheng}},\ and\ \bibinfo {author} {\bibfnamefont {Y.}~\bibnamefont {Yan}},\
  }\bibfield  {title} {\enquote {\bibinfo {title} {{Improving the efficiency of
  hierarchical equations of motion approach and application to coherent
  dynamics in Aharonov–Bohm interferometers}},}\ }\href
  {https://doi.org/10.1063/1.4914514} {\bibfield  {journal} {\bibinfo
  {journal} {The Journal of Chemical Physics}\ }\textbf {\bibinfo {volume}
  {142}},\ \bibinfo {pages} {104112} (\bibinfo {year} {2015})}\BibitemShut
  {NoStop}%
\bibitem [{\citenamefont {Shi}\ \emph {et~al.}(2018)\citenamefont {Shi},
  \citenamefont {Xu}, \citenamefont {Yan},\ and\ \citenamefont
  {Xu}}]{ShiJCP2018Tensor}%
  \BibitemOpen
  \bibfield  {author} {\bibinfo {author} {\bibfnamefont {Q.}~\bibnamefont
  {Shi}}, \bibinfo {author} {\bibfnamefont {Y.}~\bibnamefont {Xu}}, \bibinfo
  {author} {\bibfnamefont {Y.}~\bibnamefont {Yan}},\ and\ \bibinfo {author}
  {\bibfnamefont {M.}~\bibnamefont {Xu}},\ }\bibfield  {title} {\enquote
  {\bibinfo {title} {{Efficient propagation of the hierarchical equations of
  motion using the matrix product state method}},}\ }\href
  {https://doi.org/10.1063/1.5026753} {\bibfield  {journal} {\bibinfo
  {journal} {The Journal of Chemical Physics}\ }\textbf {\bibinfo {volume}
  {148}},\ \bibinfo {pages} {174102} (\bibinfo {year} {2018})}\BibitemShut
  {NoStop}%
\bibitem [{\citenamefont {Borrelli}(2019)}]{BorrelliCP2018Tensor}%
  \BibitemOpen
  \bibfield  {author} {\bibinfo {author} {\bibfnamefont {R.}~\bibnamefont
  {Borrelli}},\ }\bibfield  {title} {\enquote {\bibinfo {title} {{Density
  matrix dynamics in twin-formulation: An efficient methodology based on
  tensor-train representation of reduced equations of motion}},}\ }\href
  {https://doi.org/10.1063/1.5099416} {\bibfield  {journal} {\bibinfo
  {journal} {The Journal of Chemical Physics}\ }\textbf {\bibinfo {volume}
  {150}},\ \bibinfo {pages} {234102} (\bibinfo {year} {2019})}\BibitemShut
  {NoStop}%
\bibitem [{\citenamefont {Press}\ \emph {et~al.}(1988)\citenamefont {Press},
  \citenamefont {Vetterling}, \citenamefont {Teukolsky},\ and\ \citenamefont
  {Flannery}}]{press1988numerical}%
  \BibitemOpen
  \bibfield  {author} {\bibinfo {author} {\bibfnamefont {W.~H.}\ \bibnamefont
  {Press}}, \bibinfo {author} {\bibfnamefont {W.~T.}\ \bibnamefont
  {Vetterling}}, \bibinfo {author} {\bibfnamefont {S.~A.}\ \bibnamefont
  {Teukolsky}},\ and\ \bibinfo {author} {\bibfnamefont {B.~P.}\ \bibnamefont
  {Flannery}},\ }\href@noop {} {\emph {\bibinfo {title} {Numerical recipes}}}\
  (\bibinfo  {publisher} {Cambridge University Press},\ \bibinfo {year}
  {1988})\BibitemShut {NoStop}%
\bibitem [{\citenamefont {Strümpfer}\ and\ \citenamefont
  {Schulten}(2012)}]{doi:10.1021/ct3003833}%
  \BibitemOpen
  \bibfield  {author} {\bibinfo {author} {\bibfnamefont {J.}~\bibnamefont
  {Strümpfer}}\ and\ \bibinfo {author} {\bibfnamefont {K.}~\bibnamefont
  {Schulten}},\ }\bibfield  {title} {\enquote {\bibinfo {title} {Open quantum
  dynamics calculations with the hierarchy equations of motion on parallel
  computers},}\ }\href {https://doi.org/10.1021/ct3003833} {\bibfield
  {journal} {\bibinfo  {journal} {Journal of Chemical Theory and Computation}\
  }\textbf {\bibinfo {volume} {8}},\ \bibinfo {pages} {2808--2816} (\bibinfo
  {year} {2012})},\ \bibinfo {note} {pMID: 23105920}\BibitemShut {NoStop}%
\bibitem [{\citenamefont {Kreisbeck}, \citenamefont {Kramer},\ and\
  \citenamefont {Aspuru-Guzik}(2014)}]{doi:10.1021/ct500629s}%
  \BibitemOpen
  \bibfield  {author} {\bibinfo {author} {\bibfnamefont {C.}~\bibnamefont
  {Kreisbeck}}, \bibinfo {author} {\bibfnamefont {T.}~\bibnamefont {Kramer}},\
  and\ \bibinfo {author} {\bibfnamefont {A.}~\bibnamefont {Aspuru-Guzik}},\
  }\bibfield  {title} {\enquote {\bibinfo {title} {Scalable high-performance
  algorithm for the simulation of exciton dynamics. application to the
  light-harvesting complex ii in the presence of resonant vibrational modes},}\
  }\href {https://doi.org/10.1021/ct500629s} {\bibfield  {journal} {\bibinfo
  {journal} {Journal of Chemical Theory and Computation}\ }\textbf {\bibinfo
  {volume} {10}},\ \bibinfo {pages} {4045--4054} (\bibinfo {year} {2014})},\
  \bibinfo {note} {pMID: 26588548}\BibitemShut {NoStop}%
\bibitem [{\citenamefont {Kramer}\ \emph {et~al.}(2018)\citenamefont {Kramer},
  \citenamefont {Noack}, \citenamefont {Reimers}, \citenamefont {Reinefeld},
  \citenamefont {Rodríguez},\ and\ \citenamefont {Yin}}]{KRAMER2018262}%
  \BibitemOpen
  \bibfield  {author} {\bibinfo {author} {\bibfnamefont {T.}~\bibnamefont
  {Kramer}}, \bibinfo {author} {\bibfnamefont {M.}~\bibnamefont {Noack}},
  \bibinfo {author} {\bibfnamefont {J.~R.}\ \bibnamefont {Reimers}}, \bibinfo
  {author} {\bibfnamefont {A.}~\bibnamefont {Reinefeld}}, \bibinfo {author}
  {\bibfnamefont {M.}~\bibnamefont {Rodríguez}},\ and\ \bibinfo {author}
  {\bibfnamefont {S.}~\bibnamefont {Yin}},\ }\bibfield  {title} {\enquote
  {\bibinfo {title} {Energy flow in the photosystem i supercomplex: Comparison
  of approximative theories with \uppercase{D}m-heom},}\ }\href
  {https://doi.org/https://doi.org/10.1016/j.chemphys.2018.05.028} {\bibfield
  {journal} {\bibinfo  {journal} {Chemical Physics}\ }\textbf {\bibinfo
  {volume} {515}},\ \bibinfo {pages} {262--271} (\bibinfo {year} {2018})},\
  \bibinfo {note} {ultrafast Photoinduced Processes in Polyatomic
  Molecules:Electronic Structure, Dynamics and Spectroscopy (Dedicated to
  Wolfgang Domcke on the occasion of his 70th birthday)}\BibitemShut {NoStop}%
\bibitem [{\citenamefont {Kreisbeck}\ \emph {et~al.}(2011)\citenamefont
  {Kreisbeck}, \citenamefont {Kramer}, \citenamefont {Rodríguez},\ and\
  \citenamefont {Hein}}]{doi:10.1021/ct200126d}%
  \BibitemOpen
  \bibfield  {author} {\bibinfo {author} {\bibfnamefont {C.}~\bibnamefont
  {Kreisbeck}}, \bibinfo {author} {\bibfnamefont {T.}~\bibnamefont {Kramer}},
  \bibinfo {author} {\bibfnamefont {M.}~\bibnamefont {Rodríguez}},\ and\
  \bibinfo {author} {\bibfnamefont {B.}~\bibnamefont {Hein}},\ }\bibfield
  {title} {\enquote {\bibinfo {title} {High-performance solution of
  hierarchical equations of motion for studying energy transfer in
  light-harvesting complexes},}\ }\href {https://doi.org/10.1021/ct200126d}
  {\bibfield  {journal} {\bibinfo  {journal} {Journal of Chemical Theory and
  Computation}\ }\textbf {\bibinfo {volume} {7}},\ \bibinfo {pages}
  {2166--2174} (\bibinfo {year} {2011})},\ \bibinfo {note} {pMID:
  26606486}\BibitemShut {NoStop}%
\bibitem [{\citenamefont {Hein}\ \emph {et~al.}(2012)\citenamefont {Hein},
  \citenamefont {Kreisbeck}, \citenamefont {Kramer},\ and\ \citenamefont
  {Rodríguez}}]{Hein_2012}%
  \BibitemOpen
  \bibfield  {author} {\bibinfo {author} {\bibfnamefont {B.}~\bibnamefont
  {Hein}}, \bibinfo {author} {\bibfnamefont {C.}~\bibnamefont {Kreisbeck}},
  \bibinfo {author} {\bibfnamefont {T.}~\bibnamefont {Kramer}},\ and\ \bibinfo
  {author} {\bibfnamefont {M.}~\bibnamefont {Rodríguez}},\ }\bibfield  {title}
  {\enquote {\bibinfo {title} {Modelling of oscillations in two-dimensional
  echo-spectra of the fenna–matthews–olson complex},}\ }\href
  {https://doi.org/10.1088/1367-2630/14/2/023018} {\bibfield  {journal}
  {\bibinfo  {journal} {New Journal of Physics}\ }\textbf {\bibinfo {volume}
  {14}},\ \bibinfo {pages} {023018} (\bibinfo {year} {2012})}\BibitemShut
  {NoStop}%
\bibitem [{\citenamefont {Tsuchimoto}\ and\ \citenamefont
  {Tanimura}(2015)}]{TT15JCTC}%
  \BibitemOpen
  \bibfield  {author} {\bibinfo {author} {\bibfnamefont {M.}~\bibnamefont
  {Tsuchimoto}}\ and\ \bibinfo {author} {\bibfnamefont {Y.}~\bibnamefont
  {Tanimura}},\ }\bibfield  {title} {\enquote {\bibinfo {title} {Spins dynamics
  in a dissipative environment: Hierarchal equations of motion approach using a
  \uppercase{G}raphics \uppercase{P}rocessing \uppercase{U}nit
  (\uppercase{GPU})},}\ }\href {https://doi.org/10.1021/acs.jctc.5b00488}
  {\bibfield  {journal} {\bibinfo  {journal} {Journal of Chemical Theory and
  Computation}\ }\textbf {\bibinfo {volume} {11}},\ \bibinfo {pages}
  {3859--3865} (\bibinfo {year} {2015})}\BibitemShut {NoStop}%
\bibitem [{\citenamefont {Okumura}\ and\ \citenamefont
  {Tanimura}(1996)}]{OT96JCP}%
  \BibitemOpen
  \bibfield  {author} {\bibinfo {author} {\bibfnamefont {K.}~\bibnamefont
  {Okumura}}\ and\ \bibinfo {author} {\bibfnamefont {Y.}~\bibnamefont
  {Tanimura}},\ }\bibfield  {title} {\enquote {\bibinfo {title} {Unified
  time-path approach to the effect of anharmonicity on the molecular
  vibrational spectroscopy in solution},}\ }\href
  {https://doi.org/10.1063/1.472589} {\bibfield  {journal} {\bibinfo  {journal}
  {The Journal of Chemical Physics}\ }\textbf {\bibinfo {volume} {105}},\
  \bibinfo {pages} {7294--7309} (\bibinfo {year} {1996})}\BibitemShut {NoStop}%
\bibitem [{\citenamefont {Tanimura}\ and\ \citenamefont
  {Okumura}(1997)}]{TO97JCP}%
  \BibitemOpen
  \bibfield  {author} {\bibinfo {author} {\bibfnamefont {Y.}~\bibnamefont
  {Tanimura}}\ and\ \bibinfo {author} {\bibfnamefont {K.}~\bibnamefont
  {Okumura}},\ }\bibfield  {title} {\enquote {\bibinfo {title} {First-, third-,
  and fifth-order resonant spectroscopy of an anharmonic displaced oscillators
  system in the condensed phase},}\ }\href {https://doi.org/10.1063/1.473099}
  {\bibfield  {journal} {\bibinfo  {journal} {The Journal of Chemical Physics}\
  }\textbf {\bibinfo {volume} {106}},\ \bibinfo {pages} {2078--2095} (\bibinfo
  {year} {1997})}\BibitemShut {NoStop}%
\end{thebibliography}%

\end{document}